\documentclass[prd,aps,nofootinbib,floatfix,11pt]{revtex4}

\usepackage{amsmath,graphicx,epsfig,amssymb,dsfont,mathtools}
\usepackage[usenames]{color}
\usepackage{ulem} 
\usepackage{bigstrut}
\usepackage{slashed}
\usepackage{multirow}
\usepackage{subfigure}

\allowdisplaybreaks

\begin{document}

\title{Weak decays of triply heavy baryons in the light-front approach}
 
\author{Zhen-Xing Zhao}\email{zhaozx19@imu.edu.cn}
\author{Qi Yang}
\affiliation{School of Physical Science and Technology, Inner Mongolia University, Hohhot 010021, China}
 
\begin{abstract}
In this work, we investigate the weak decays of ground-state triply
heavy baryons. We first obtain the form factors using the light-front
quark model in the three-quark picture, and then apply them to arrive
at some phenomenological predictions, including the decay widths of
semileptonic decays and nonleptonic decays. Our results are expected
to be helpful for the experimental search for triply heavy baryons.
\end{abstract}

\maketitle

\section{Introduction}

Since LHCb reported the discovery of the doubly charmed baryon $\Xi_{cc}^{++}$
in the final state $\Lambda_{c}^{+}K^{-}\pi^{+}\pi^{+}$~\cite{LHCb:2017iph},
the experimentalists have made great progress in the field of doubly
heavy baryons (DHBs) -- They measured the lifetime of $\Xi_{cc}^{++}$~\cite{LHCb:2018zpl},
and then found two other new channels $\Xi_{cc}^{++}\to\Xi_{c}^{+}\pi^{+}$
\cite{LHCb:2018pcs} and $\Xi_{cc}^{++}\to\Xi_{c}^{\prime+}\pi^{+}$~\cite{LHCb:2022rpd}.
Surrounded by optimism, people tend to expect that triply heavy baryons
(THBs) may be discovered in the near future. The theoretical research
on THBs are highly demanded. 

Some efforts have been made in this direction. Ref.~\cite{Wang:2020avt}
obtained the masses of the ground-state THBs using QCD sum rules.
In Ref.~\cite{Huang:2021jxt}, the authors performed an analysis
on the weak decays with the help of SU(3) flavor symmetry. In recent
works~\cite{Wang:2022ias,Lu:2024bqw}, the authors investigated
the weak decays of $\Omega_{ccc}^{++}$ and $\Omega_{bbb}^{-}$. More
works can be found in Refs.~\cite{GomshiNobary:2004mq,GomshiNobary:2005ur,Jia:2006gw,GomshiNobary:2006tzy,Martynenko:2007je,Patel:2008mv,Zhang:2009re,Zheng:2010zzc,Chen:2011mb,Flynn:2011gf,Llanes-Estrada:2011gwu,Wang:2011ae,Albertus:2012isp,Aliev:2012tt,Aliev:2014lxa,Azizi:2014jxa,Vijande:2015faa,Thakkar:2016cdn,Shah:2017jkr,MoosaviNejad:2017rvi,Rai:2017hue,Bhavsar:2018tad,Wang:2018utj,Shah:2019jxp,Yang:2019lsg,Delpasand:2019xpk,Alomayrah:2020qyw,Tazimi:2021ywr,Ghasemi:2021wqo,Mutuk:2021zes,Faustov:2021qqf}.
However, there is still a lack of one comprehensive quantitative analysis
on the weak decays of THBs. This work and the forthcoming ones aim
to fill this gap.

In this work, we only consider S-wave THBs, which can only be spin-3/2
or spin-1/2. We have, for the former, $\Omega_{ccc}^{++}$, $\Omega_{bcc}^{\prime+}$,
$\Omega_{bbc}^{\prime0}$ and $\Omega_{bbb}^{-}$, while for the latter,
$\Omega_{bcc}^{+}$ and $\Omega_{bbc}^{0}$. \footnote{Readers may have noticed our naming principle here: keep the symbols
of the ground states as simple as possible. The same goes for DHBs. } Their flavor wavefunctions are given as:
\begin{align}
\Omega_{QQQ} & =QQQ,\nonumber \\
\Omega_{QQQ^{\prime}}^{\prime} & =\frac{1}{\sqrt{3}}(Q^{\prime}QQ+QQ^{\prime}Q+QQQ^{\prime}),\nonumber \\
\Omega_{QQQ^{\prime}} & =QQQ^{\prime},
\end{align}
where $Q,Q^{\prime}=c/b$. It can be seen that, for the spin-3/2 THBs,
the three heavy quarks are on an equal footing, while for the spin-1/2
THBs, the two identical heavy quarks are usually considered to form
an axial-vector diquark. The spin-3/2 $\Omega_{QQQ^{\prime}}^{\prime}$
and spin-1/2 $\Omega_{QQQ^{\prime}}$ have the same quark components,
while the latter is usually considered as the ground state. As can
be seen in Ref. \cite{Sun:2023noo}, the flavor mixing between baryons
with the same quark components, is ubiquitous. The dynamics mechanism
comes from the gluon exchange between quarks inside baryons. However,
the mixing is small when a heavy quark serves as a source of color
charge. In a word, the mass eigenstates discovered in experiments
are the flavor mixing states of $\Omega_{QQQ^{\prime}}^{\prime}$
and $\Omega_{QQQ^{\prime}}$, but this mixing can be neglected. The
experimentalists should prioritize searching for the ground-state
$\Omega_{ccc}^{++}$, $\Omega_{bbb}^{-}$, and $\Omega_{bcc}^{+}$,
$\Omega_{bbc}^{0}$.

Therefore, in this work, we will investigate the weak decays of THBs
for the $3/2\to1/2$ case and $1/2\to1/2$ case, and the method of
light-front quark model (LFQM) will be adopted. LFQM has been widely
used to study the properties of mesons~\cite{Jaus:1999zv,Jaus:1989au,Jaus:1991cy,Cheng:1996if,Cheng:2003sm,Cheng:2004yj,Ke:2009ed,Ke:2009mn,Cheng:2009ms,Lu:2007sg,Wang:2007sxa,Wang:2008xt,Wang:2008ci,Wang:2009mi,Chen:2009qk,Li:2010bb,Verma:2011yw,Shi:2016gqt,Chang:2018aut,Chang:2018zjq,Chang:2019mmh,Chang:2019obq,Chang:2020wvs,Chen:2021ywv},
and has achieved great success. In recent years, LFQM is further applied
to the baryon sector: in Refs.~\cite{Ke:2007tg,Wei:2009np,Ke:2012wa,Hu:2020mxk,Hsiao:2020gtc},
the diquark picture is adopted, while in Refs.~\cite{Ke:2019smy,Ke:2019lcf,Ke:2021pxk,Geng:2020fng,Geng:2020gjh,Geng:2021nkl,Geng:2022xpn,Zhao:2023yuk,Xing:2023jnr},
the three-quark picture is used. In the diquark picture, the two loosely-bound
spectator quarks are viewed as a whole to form so-called ``diquark'',
whose spin-parity can only be $0^{+}$ or $1^{+}$ for a ground-state
baryon. The diquark picture has some inevitable flaws. Firstly,
a diquark is arbitrarily designated for the convenience of research.
Secondly, the diquark picture includes parameters such as the diquark
mass that are difficult to determine. Lastly, the overlap factor,
which is actually the inner product of the flavor-spin wavefunctions
of the initial and final baryons, is not easy to obtain in the diquark
picture. In this work, we will adopt the three-quark picture. 

Specifically, in this work, we will consider the following $3/2\to1/2$
processes
\begin{align}
\Omega_{ccc}^{++}(ccc) & \to\Xi_{cc}^{+}(dcc)/\Omega_{cc}^{+}(scc),\nonumber \\
\Omega_{bcc}^{\prime+}(ccb) & \to\Xi_{bc}^{\prime0}(dcb)/\Omega_{bc}^{\prime0}(scb),\nonumber \\
\Omega_{bbc}^{\prime0}(cbb) & \to\Xi_{bb}^{-}(dbb)/\Omega_{bb}^{-}(sbb),\nonumber \\
\Omega_{bcc}^{\prime+}(bcc) & \to\Xi_{cc}^{++}(ucc),\nonumber \\
\Omega_{bbc}^{\prime0}(bbc) & \to\Xi_{bc}^{\prime+}(ubc),\nonumber \\
\Omega_{bbb}^{-}(bbb) & \to\Xi_{bb}^{0}(ubb).
\end{align}
and the following $1/2\to1/2$ processes
\begin{align}
\Omega_{bcc}^{+}(ccb) & \to\Xi_{bc}^{(\prime)0}(dcb)/\Omega_{bc}^{(\prime)0}(scb),\nonumber \\
\Omega_{bbc}^{0}(cbb) & \to\Xi_{bb}^{-}(dbb)/\Omega_{bb}^{-}(sbb),\nonumber \\
\Omega_{bcc}^{+}(bcc) & \to\Xi_{cc}^{++}(ucc),\nonumber \\
\Omega_{bbc}^{0}(bbc) & \to\Xi_{bc}^{(\prime)+}(ubc),
\end{align}
where for the final-state DHBs, we have denoted, for example, 
\begin{align}
\Xi_{bc}^{0} & =\frac{1}{2}(bc-cb)d,\nonumber \\
\Xi_{bc}^{\prime0} & =\frac{1}{2}(bc+cb)d.
\end{align}
For $\Xi_{bc}^{(\prime)0}$, the two heavy quarks are usually considered
as a scalar (axial-vector) diquark, while for $\Xi_{QQ}$, the two
identical heavy quarks are considered to form an axial-vector diquark.
For more details on the classification of ground-state DHBs, one can
refer to Ref.~\cite{Wang:2017mqp}. The $1/2\to1/2$ processes can
be further divided into two categories -- according to whether the
system of two spectator quarks is a scalar or axial vector. However,
for the $3/2\to1/2$ processes, the system of the two spectator quarks
can only be an axial vector. Therefore, we will consider the following
three typical processes: 
\begin{itemize}
\item $\Omega_{bcc}^{\prime+}(ccb)\to\Omega_{bc}^{\prime0}(scb)$ for the
$3/2\to1/2$ case, 
\item $\Omega_{bcc}^{+}(ccb)\to\Omega_{bc}^{0}(scb)$ for the $1/2\to1/2$
case with a scalar spectator diquark,
\item $\Omega_{bcc}^{+}(ccb)\to\Omega_{bc}^{\prime0}(scb)$ for the $1/2\to1/2$
case with an axial-vector spectator diquark.
\end{itemize}

The rest of this paper is arranged as follows. In Sec.~II, the framework
of light-front approach under the three quark-picture is briefly introduced.
We will discuss the transition matrix elements for three typical processes
and extract the corresponding form factors. The non-trivial overlap
factors in flavor-spin space are also shown. In Sec.~III, the numerical
results for the form factors are presented, and then used to give
some phenomenological predictions. We conclude this paper in the last
section.

\section{Form factors in the light-front approach}

\subsection{The light-front approach}

In Ref. \cite{Zhao:2023yuk}, we reviewed the light-front quark model
of baryons under the three-quark picture and provided constructive
proofs for the spin wavefunctions of S-wave baryons. Some of the main
results are listed below. 

In the framework of light-front approach under the three-quark picture,
a baryon state is expressed as
\begin{eqnarray}
|{\cal B}(P,S,S_{z})\rangle & = & \int\{d^{3}\tilde{p}_{1}\}\{d^{3}\tilde{p}_{2}\}\{d^{3}\tilde{p}_{3}\}\ 2(2\pi)^{3}\delta^{3}(\tilde{P}-\tilde{p}_{1}-\tilde{p}_{2}-\tilde{p}_{3})\frac{1}{\sqrt{P^{+}}}\nonumber \\
 &  & \times\sum_{\lambda_{1},\lambda_{2},\lambda_{3}}\Psi^{SS_{z}}(\tilde{p}_{1},\tilde{p}_{2},\tilde{p}_{3},\lambda_{1},\lambda_{2},\lambda_{3})C^{ijk}|q_{1}^{i}(p_{1},\lambda_{1})q_{2}^{j}(p_{2},\lambda_{2})q_{3}^{k}(p_{3},\lambda_{3})\rangle,\label{eq:state_vector}
\end{eqnarray}
where $p_{i}$ ($\lambda_{i}$) is the light-front momentum (helicity)
of the quark $q_{i}$, the color wavefunction $C^{ijk}=\epsilon^{ijk}/\sqrt{6}$,
and the flavor-spin and momentum wavefunctions are contained in $\Psi^{SS_{z}}$.
The light-front momentum is decomposed into $p_{i}=(p_{i}^{-},p_{i}^{+},p_{i\perp})$
with $p_{i}^{\pm}\equiv p_{i}^{0}\pm p_{i}^{3}$ and $p_{i\perp}=(p_{i}^{1},p_{i}^{2})$,
and the quarks are assumed to be on mass shell:
\begin{equation}
p_{i}^{-}=\frac{m_{i}^{2}+p_{i\perp}^{2}}{p_{i}^{+}}.
\end{equation}
In Eq. (\ref{eq:state_vector}),
\begin{equation}
\tilde{p}_{i}=(p_{i}^{+},p_{i\perp}),\quad\{d^{3}\tilde{p}_{i}\}=\frac{dp_{i}^{+}d^{2}p_{i\perp}}{2(2\pi)^{3}}.
\end{equation}
The intrinsic variables $(x_{i},k_{i\perp})$ are introduced through
\begin{align}
 & p_{i}^{+}=x_{i}P^{+},\quad p_{i\perp}=x_{i}P_{\perp}+k_{i\perp},\nonumber \\
 & \sum_{i=1}^{3}x_{i}=1,\quad\sum_{i=1}^{3}k_{i\perp}=0,
\end{align}
where $x_{i}$ is the light-front momentum fraction constrained by
$0\le x_{i}\le1$. Define $\bar{P}^{\mu}\equiv p_{1}+p_{2}+p_{3}$
and $M_{0}^{2}\equiv\bar{P}^{2}$, and it can be shown that
\begin{equation}
M_{0}^{2}=\frac{k_{1\perp}^{2}+m_{1}^{2}}{x_{1}}+\frac{k_{2\perp}^{2}+m_{2}^{2}}{x_{2}}+\frac{k_{3\perp}^{2}+m_{3}^{2}}{x_{3}}.
\end{equation}

The internal momenta are defined by
\begin{equation}
k_{i}=(k_{i}^{-},k_{i}^{+},k_{i\perp})=(e_{i}-k_{iz},e_{i}+k_{iz},k_{i\perp})=(\frac{m_{i}^{2}+k_{i\perp}^{2}}{x_{i}M_{0}},x_{i}M_{0},k_{i\perp}),
\end{equation}
then it is easy to verify:
\begin{align}
e_{i} & =\frac{x_{i}M_{0}}{2}+\frac{m_{i}^{2}+k_{i\perp}^{2}}{2x_{i}M_{0}},\nonumber \\
k_{iz} & =\frac{x_{i}M_{0}}{2}-\frac{m_{i}^{2}+k_{i\perp}^{2}}{2x_{i}M_{0}},
\end{align}
where $e_{i}$ is the energy of the quark $q_{i}$ in the rest frame
of $\bar{P}$.

For S-wave baryons, three typical wavefunctions are given as follows: 
\begin{itemize}
\item $\Lambda_{Q}$-type:
\begin{align}
\Psi_{0}^{S=\frac{1}{2},S_{z}}(\tilde{p}_{i},\lambda_{i})= & A_{0}\bar{u}(p_{3},\lambda_{3})(\bar{\slashed P}+M_{0})(-\gamma_{5})C\bar{u}^{T}(p_{2},\lambda_{2})\nonumber \\
 & \times\bar{u}(p_{1},\lambda_{1})u(\bar{P},S_{z})\Phi(x_{i},k_{i\perp}),
\end{align}
\item $\Sigma_{Q}$-type:
\begin{align}
\Psi_{1}^{S=\frac{1}{2},S_{z}}(\tilde{p}_{i},\lambda_{i})= & A_{1}\bar{u}(p_{3},\lambda_{3})(\bar{\slashed P}+M_{0})(\gamma^{\mu}-v^{\mu})C\bar{u}^{T}(p_{2},\lambda_{2})\nonumber \\
 & \times\bar{u}(p_{1},\lambda_{1})(\frac{1}{\sqrt{3}}\gamma_{\mu}\gamma_{5})u(\bar{P},S_{z})\Phi(x_{i},k_{i\perp}),
\end{align}
\item $\Sigma_{Q}^{*}$-type:
\begin{align}
\Psi_{1}^{S=\frac{3}{2},S_{z}}(\tilde{p}_{i},\lambda_{i})= & A_{1}^{\prime}\bar{u}(p_{3},\lambda_{3})(\bar{\slashed P}+M_{0})(\gamma^{\mu}-v^{\mu})C\bar{u}^{T}(p_{2},\lambda_{2})\nonumber \\
 & \times\bar{u}(p_{1},\lambda_{1})u_{\mu}(\bar{P},S_{z})\Phi(x_{i},k_{i\perp}),
\end{align}
\end{itemize}
where quark $1$ is $Q=b/c$ while quarks $2,3$ are $u,d$ or $d,u$,
$v^{\mu}\equiv\bar{P}^{\mu}/M_{0}$, $\Phi$ is the momentum wavefunction,
and the normalization factors
\begin{equation}
A_{0}=A_{1}=A_{1}^{\prime}=\frac{1}{4\sqrt{M_{0}^{3}(e_{1}+m_{1})(e_{2}+m_{2})(e_{3}+m_{3})}}.\label{eq:normalization}
\end{equation}

More discussion is needed regarding the three types of spin wavefunctions.
Respectively denote the spin wavefunctions of $\Lambda_{Q}$, $\Sigma_{Q}$
and $\Sigma_{Q}^{*}$ as
\begin{align}
\psi_{0}(321) & \equiv\bar{u}(p_{3},\lambda_{3})(\bar{\slashed P}+M_{0})(-\gamma_{5})C\bar{u}^{T}(p_{2},\lambda_{2})\bar{u}(p_{1},\lambda_{1})u(\bar{P},S_{z}),\nonumber \\
\psi_{1}(321) & \equiv\bar{u}(p_{3},\lambda_{3})(\bar{\slashed P}+M_{0})(\gamma^{\mu}-v^{\mu})C\bar{u}^{T}(p_{2},\lambda_{2})\bar{u}(p_{1},\lambda_{1})(\frac{1}{\sqrt{3}}\gamma_{\mu}\gamma_{5})u(\bar{P},S_{z}),\nonumber \\
\psi_{1\mu}(321) & \equiv\bar{u}(p_{3},\lambda_{3})(\bar{\slashed P}+M_{0})(\gamma^{\mu}-v^{\mu})C\bar{u}^{T}(p_{2},\lambda_{2})\bar{u}(p_{1},\lambda_{1})u_{\mu}(\bar{P},S_{z}).
\end{align}
In $\psi_{0}(321)$, quarks 3 and 2 are usually considered to form
a scalar diquark, while in $\psi_{1}(321)$ and $\psi_{1\mu}(321)$,
they are considered to form an axial-vector diquark. It can be shown
that $\psi_{0,1,1\mu}(321)$ have the same normalization factor $A_{0}=A_{1}=A_{1}^{\prime}$.
As expected, a scalar diquark and an axis-vector diquark are orthogonal,
so $\psi_{0}(321)$ is orthogonal to $\psi_{1}(321)$ and $\psi_{1\mu}(321)$.
Therefore, under the valence quark approximation, the amplitude of
$\Lambda_{b}\to\Sigma_{c}$ is zero. One can easily check that, 
\begin{align}
\psi_{0}(231) & =-\psi_{0}(321),\nonumber \\
\psi_{1}(231) & =+\psi_{1}(321),
\end{align}
and at least in the sense of spin coupling
\begin{equation}
\left(\begin{array}{c}
\psi_{0}(312)\\
\psi_{1}(312)
\end{array}\right)=\left(\begin{array}{cc}
\frac{1}{2} & -\frac{\sqrt{3}}{2}\\
-\frac{\sqrt{3}}{2} & -\frac{1}{2}
\end{array}\right)\left(\begin{array}{c}
\psi_{0}(321)\\
\psi_{1}(321)
\end{array}\right).\label{eq:T_matrix}
\end{equation}
Also at least in the sense of spin coupling, it can be checked that
\begin{equation}
\psi_{1\mu}(321)=\psi_{1\mu}(213)=\psi_{1\mu}(132)=\cdots,
\end{equation}
which is as expected because the three quarks in $\psi_{1\mu}(321)$
are on an equal footing.

For the momentum wavefunction, we will adopt the following form
\begin{equation}
\Phi(x_{i},k_{i\perp})=\sqrt{\frac{e_{1}e_{2}e_{3}}{x_{1}x_{2}x_{3}M_{0}}}\times3^{3/4}\times16\left(\frac{\pi}{\beta^{2}}\right)^{3/2}\exp\left(-\frac{\vec{k}_{1}^{2}+\vec{k}_{2}^{2}+\vec{k}_{3}^{2}}{2\beta^{2}}\right),\label{eq:momentum_wf}
\end{equation}
where $\vec{k}_{i}\equiv(k_{i\perp},k_{iz})$, $\beta$ is the shape
parameter that characterizes the momentum distribution inside a baryon.
In Eq. (\ref{eq:momentum_wf}), we have put the three momenta $\vec{k}_{1,2,3}$
on an equal footing, and assumed that their distributions all follow
a Gaussian distribution. A similar assumption has also been adopted
in Ref. \cite{Xing:2023jnr}. The verification of normalization for
Eq. (\ref{eq:momentum_wf}) is included in Appendix \ref{app:normalization_momentum_wf}.

Some comments on the momentum wavefunction are in order. In the literature,
the following momentum wavefunction is often used (see, for example,
Refs. \cite{Ke:2019smy,Zhao:2023yuk,Lu:2024bqw}):
\begin{equation}
\Phi(x_{i},k_{i\perp})=\sqrt{\frac{e_{1}e_{2}e_{3}}{x_{1}x_{2}x_{3}M_{0}}}\varphi(\vec{k}_{1},\beta_{1})\varphi(\frac{\vec{k}_{2}-\vec{k}_{3}}{2},\beta_{23})\label{eq:momentum_wf_old}
\end{equation}
where
\begin{equation}
\varphi(\vec{k},\beta)=4\left(\frac{\pi}{\beta^{2}}\right)^{3/4}\exp\left(-\frac{\vec{k}^{2}}{2\beta^{2}}\right),\label{eq:gaussion}
\end{equation}
and the shape parameters $\beta_{1}$ and $\beta_{23}$ respectively
characterize the momentum distribution between quark 1 and the system
of quark 2 and quark 3, and the momentum distribution between quark
2 and quark 3. One can see from Eq. (\ref{eq:momentum_wf_old}) that,
the three quarks are not equally treated, while its form should be
suitable for discussing the internal structure of excited states of
baryons.

The shape parameter $\beta$ in Eq. (\ref{eq:momentum_wf}) can be
viewed as the most important parameter in the light-front approach.
In Ref. \cite{Zhao:2023yuk}, we proposed using the pole residue to
determine this parameter. In the following, we provide some details
for determining the shape parameter for a spin-3/2 baryon. For the
case of a spin-1/2 baryon, one can refer to our previous work \cite{Zhao:2023yuk}. 

\subsection{The shape parameter}

To be specific, we will take the spin-3/2 $\Omega_{bcc}^{\prime+}$
as an example to illustrate how to determine the shape parameter. 

On the one hand, the pole residue of a spin-3/2 baryon is defined
as:
\begin{equation}
\langle0|J_{\mu}^{bcc}|\Omega_{bcc}^{\prime+}(P,S_{z})\rangle=\lambda_{+}u_{\mu}(P,S_{z}),\label{eq:pole_residue_def}
\end{equation}
where the interpolating current of $\Omega_{bcc}^{\prime+}$ can be
given by \cite{Wang:2020avt}
\begin{equation}
J_{\mu}^{bcc}=\epsilon_{ijk}(c_{i}^{T}C\gamma_{\mu}c_{j})b_{k}
\end{equation}
with $i,j,k$ the color indices.

On the other hand, the matrix element $\langle0|J_{\mu}^{bcc}|\Omega_{bcc}^{\prime+}\rangle$
can be calculated in LFQM
\begin{align}
\langle0|J_{\mu}^{bcc}|\Omega_{bcc}^{\prime+}\rangle= & \int\frac{dx_{2}d^{2}k_{2\perp}}{2(2\pi)^{3}}\frac{dx_{3}d^{2}k_{3\perp}}{2(2\pi)^{3}}\frac{A}{\sqrt{x_{1}x_{2}x_{3}}}\Phi(x_{i},k_{i\perp})\times\sqrt{6}\times\frac{1}{\sqrt{2}}\times2\nonumber \\
 & \times{\rm Tr}[C\gamma_{\mu}(\slashed p_{3}+m_{3})(\bar{\slashed P}+M_{0})(\gamma^{\rho}-v^{\rho})C(\slashed p_{2}+m_{2})^{T}]\nonumber \\
 & \times(\slashed p_{1}+m_{1})u_{\rho}(\bar{P},S_{z}),\label{eq:pole_residue_LFQM}
\end{align}
where $\sqrt{6}$ comes from the color space, $1/\sqrt{2}$ comes
from the normalization of the baryon state, $2$ comes from two equivalent
contractions, and the normalization factor $A$ can be found in Eq.
(\ref{eq:normalization}).

Multiplying Eqs. (\ref{eq:pole_residue_def}) and (\ref{eq:pole_residue_LFQM})
with $\sum_{S_{z}}\bar{u}^{\mu}(P,S_{z})\gamma^{+}$ from the left,
also noting that $\gamma^{+}u_{\mu}(P)=\gamma^{+}u_{\mu}(\bar{P})$
\cite{Chua:2018lfa}, one can obtain the expression of the pole residue
in LFQM. Then, use the pole residue obtained from other theoretical
methods, such as QCD sum rules, as input to determine the shape parameter. 

Once the shape parameter is determined, one can proceed to calculate
the weak decay form factors. 

\subsection{Form factors of the $3/2\to1/2$ case}

In this subsection, we will take $\Omega_{bcc}^{\prime+}\to\Omega_{bc}^{\prime0}$
as an example to illustrate how to extract the form factors of the
$3/2\to1/2$ case. This is a $c\to s$ process, and the $b$ quark
and one $c$ quark are spectators. For the initial state, whether
in flavor space or spin space, the three quarks are considered to
be on an equal footing. Especially, one can consider the two spectator
quarks as an axial-vector diquark. For the final state, the two spectator
quarks can only be an axial-vector diquark, which has been
mentioned above. 

One the one hand, the weak decay matrix element can be parameterized
in terms of form factors
\begin{align}
\langle\Omega_{bc}^{\prime0}(P^{\prime},S_{z}^{\prime})|\bar{s}\gamma^{\mu}c|\Omega_{bcc}^{\prime+}(P,S_{z})\rangle= & \bar{u}(P^{\prime},S_{z}^{\prime})\Bigg[\gamma^{\mu}\frac{P^{\prime\alpha}}{M^{\prime}}f_{1}^{V}(q^{2})+\frac{P^{\mu}P^{\prime\alpha}}{MM^{\prime}}f_{2}^{V}(q^{2})\nonumber \\
 & \qquad\qquad+\frac{P^{\prime\mu}P^{\prime\alpha}}{M^{\prime2}}f_{3}^{V}(q^{2})+g^{\mu\alpha}f_{4}^{V}(q^{2})\Bigg]\gamma_{5}u_{\alpha}(P,S_{z}),\label{eq:parameterization_31_V}\\
\langle\Omega_{bc}^{\prime0}(P^{\prime},S_{z}^{\prime})|\bar{s}\gamma^{\mu}\gamma_{5}c|\Omega_{bcc}^{\prime+}(P,S_{z})\rangle= & \bar{u}(P^{\prime},S_{z}^{\prime})\Bigg[\gamma^{\mu}\frac{P^{\prime\alpha}}{M^{\prime}}f_{1}^{A}(q^{2})+\frac{P^{\mu}P^{\prime\alpha}}{MM^{\prime}}f_{2}^{A}(q^{2})\nonumber \\
 & \qquad\qquad+\frac{P^{\prime\mu}P^{\prime\alpha}}{M^{\prime2}}f_{3}^{A}(q^{2})+g^{\mu\alpha}f_{4}^{A}(q^{2})\Bigg]u_{\alpha}(P,S_{z}),\label{eq:parameterization_31_A}
\end{align}
where $q=P-P^{\prime}$, $f_{i}^{V,A}$ are the form factors, and
$M^{(\prime)}$ is the mass of the initial (final) baryon.

On the other hand, the matrix element can also be calculated in LFQM
\begin{align}
 & \langle\Omega_{bc}^{\prime0}(P^{\prime},S_{z}^{\prime})|\bar{s}\gamma^{\mu}(1-\gamma_{5})c|\Omega_{bcc}^{\prime+}(P,S_{z})\rangle\nonumber \\
= & \int\{d^{3}\tilde{p}_{2}\}\{d^{3}\tilde{p}_{3}\}\frac{A^{\prime}A}{\sqrt{p_{1}^{\prime+}p_{1}^{+}P^{\prime+}P^{+}}}\Phi^{\prime*}(x_{i}^{\prime},k_{i\perp}^{\prime})\Phi(x_{i},k_{i\perp})\times\frac{1}{\sqrt{2}}\times2\nonumber \\
 & \times{\rm Tr}[(\bar{\slashed P}+M_{0})(\gamma^{\rho}-v^{\rho})C(\slashed p_{2}+m_{2})^{T}C(\gamma^{\sigma}-v^{\prime\sigma})(\bar{\slashed P^{\prime}}+M_{0}^{\prime})(\slashed p_{3}+m_{3})]\nonumber\\
 & \times\bar{u}(\bar{P}^{\prime},S_{z}^{\prime})(\frac{1}{\sqrt{3}}\gamma_{\sigma}\gamma_{5})(\slashed p_{1}^{\prime}+m_{1}^{\prime})\gamma^{\mu}(1-\gamma_{5})(\slashed p_{1}+m_{1})u_{\rho}(\bar{P},S_{z}),\label{eq:me_31}
\end{align}
where $v^{\rho}\equiv\bar{P}^{\rho}/M_{0}$, $v^{\prime\sigma}\equiv\bar{P}^{\prime\sigma}/M_{0}^{\prime}$,
and the two factors $1/\sqrt{2}$ and $2$ respectively come from
the normalization of the initial state and two equivalent contractions.

The form factors $f_{i}^{V}$ can be extracted in the following way.
Respectively multiply Eq.~(\ref{eq:parameterization_31_V}) by $\sum_{S_{z},S_{z}^{\prime}}\bar{u}^{\beta}(P,S_{z})(\Gamma_{5})_{\mu\beta}u(P^{\prime},S_{z}^{\prime})$
with $(\Gamma_{5})_{\mu\beta}\in\{\gamma_{\mu}P_{\beta}^{\prime},P_{\mu}P_{\beta}^{\prime},P_{\mu}^{\prime}P_{\beta}^{\prime},g_{\mu\beta}\}\gamma_{5}$
to arrive at one set of expressions. Do the same thing for the
vector-current part of Eq.~(\ref{eq:me_31}) to arrive at the other set
of expressions, and at this point, we need to take the approximation
$P^{(\prime)}\to\bar{P}^{(\prime)}$ within the integral. Equate the
two sets of expressions to extract the form factors $f_{i}^{V}$.
Similarly, one can also extract the form factors $f_{i}^{A}$.

\subsection{Form factors of the $1/2\to1/2$ case with a scalar spectator diquark}

\label{subsec:ff_11_S}

In this subsection, we will take $\Omega_{bcc}^{+}\to\Omega_{bc}^{0}$
as an example to illustrate how to extract the form factors of the
$1/2\to1/2$ case with a scalar spectator diquark. This is also a
$c\to s$ process, and the $b$ quark and one $c$ quark are spectators.
For the initial state, the two charm quarks are considered to form
an axial-vector diquark, while for the final state, the two heavy
quarks are considered to form a scalar diquark. 

One the one hand, the weak decay matrix element can be parameterized
in terms of form factors
\begin{align}
\langle\Omega_{bc}^{0}(P^{\prime},S_{z}^{\prime})|\bar{s}\gamma^{\mu}c|\Omega_{bcc}^{+}(P,S_{z})\rangle= & \bar{u}(P^{\prime},S_{z}^{\prime})\Bigg[\gamma^{\mu}f_{1}(q^{2})+\frac{i\sigma^{\mu\nu}q_{\nu}}{M}f_{2}(q^{2})+\frac{q^{\mu}}{M}f_{3}(q^{2})\Bigg]u(P,S_{z}),\label{eq:parameterization_11_V}\\
\langle\Omega_{bc}^{0}(P^{\prime},S_{z}^{\prime})|\bar{s}\gamma^{\mu}\gamma_{5}c|\Omega_{bcc}^{+}(P,S_{z})\rangle= & \bar{u}(P^{\prime},S_{z}^{\prime})\Bigg[\gamma^{\mu}g_{1}(q^{2})+\frac{i\sigma^{\mu\nu}q_{\nu}}{M}g_{2}(q^{2})+\frac{q^{\mu}}{M}g_{3}(q^{2})\Bigg]\gamma_{5}u(P,S_{z}),\label{eq:parameterization_11_A}
\end{align}
where $q=P-P^{\prime}$, $f_{i},g_{i}$ are the form factors, and
$M$ is the mass of the initial baryon.

On the other hand, the matrix element can also be calculated in LFQM
in two different ways.

\paragraph*{(1) The direct way}

Doing the contraction directly, one can obtain
\begin{align}
 & \langle\Omega_{bc}^{0}(P^{\prime},S_{z}^{\prime})|\bar{s}\gamma^{\mu}(1-\gamma_{5})c|\Omega_{bcc}^{+}(P,S_{z})\rangle\nonumber \\
= & \int\{d^{3}\tilde{p}_{2}\}\{d^{3}\tilde{p}_{3}\}\frac{A^{\prime}A}{\sqrt{p_{1}^{\prime+}p_{1}^{+}P^{\prime+}P^{+}}}\Phi^{\prime*}(x_{i}^{\prime},k_{i\perp}^{\prime})\Phi(x_{i},k_{i\perp})\times(-1)\times\frac{1}{\sqrt{2}}\times2\nonumber \\
 & \times\bar{u}(\bar{P}^{\prime},S_{z}^{\prime})(\slashed p_{1}^{\prime}+m_{1}^{\prime})\gamma^{\mu}(1-\gamma_{5})(\slashed p_{1}+m_{1})(\bar{\slashed P}+M_{0})(\gamma^{\rho}-v^{\rho})\nonumber \\
 & \times C(\slashed p_{3}+m_{3})^{T}C\gamma_{5}(\bar{\slashed P^{\prime}}+M_{0}^{\prime})(\slashed p_{2}+m_{2})(\frac{1}{\sqrt{3}}\gamma_{\rho}\gamma_{5})u(\bar{P},S_{z}),\label{eq:me_11_S_1}
\end{align}
where the four quarks can be numbered as $(1,1^{\prime},2,3)=(c^{1},s,b,c^{2})$
with $c^{i}$ the $i$-th charm quark, and the factor $-1$ comes
from the exchange of the $b$ quark and the $c$ quark in the final
state.

\paragraph*{(2) The indirect way}

The spin wavefunction of the initial state can be written as $\psi_{1}(ccb)$,
where the two charm quarks are considered as an axial-vector diquark.
Expand $\psi_{1}(ccb)$ in terms of a new diquark basis $\psi_{0,1}(cbc)$
\begin{equation}
\psi_{1}(ccb)=(-\frac{\sqrt{3}}{2})\psi_{0}(cbc)+(-\frac{1}{2})\psi_{1}(cbc),\label{eq:ccb_2_cbc}
\end{equation}
see Eq. (\ref{eq:T_matrix}), while the spin wavefunction of the final
state is
\begin{equation}
\psi_{0}(cbs),
\end{equation}
then one can read the factor $-\sqrt{3}/2$. It turns out that
\begin{align}
 & \langle\Omega_{bc}^{0}(P^{\prime},S_{z}^{\prime})|\bar{s}\gamma^{\mu}(1-\gamma_{5})c|\Omega_{bcc}^{+}(P,S_{z})\rangle\nonumber \\
= & \int\{d^{3}\tilde{p}_{2}\}\{d^{3}\tilde{p}_{3}\}\frac{A^{\prime}A}{\sqrt{p_{1}^{\prime+}p_{1}^{+}P^{\prime+}P^{+}}}\Phi^{\prime*}(x_{i}^{\prime},k_{i\perp}^{\prime})\Phi(x_{i},k_{i\perp})\times(-\frac{\sqrt{3}}{2})\times\frac{1}{\sqrt{2}}\times2\nonumber \\
 & \times{\rm Tr}[(\bar{\slashed P}+M_{0})(-\gamma_{5})C(\slashed p_{2}+m_{2})^{T}C\gamma_{5}(\bar{\slashed P^{\prime}}+M_{0}^{\prime})(\slashed p_{3}+m_{3})]\nonumber \\
 & \times\bar{u}(\bar{P}^{\prime},S_{z}^{\prime})(\slashed p_{1}^{\prime}+m_{1}^{\prime})\gamma^{\mu}(1-\gamma_{5})(\slashed p_{1}+m_{1})u(\bar{P},S_{z}).\label{eq:me_11_S_2}
\end{align}

One can check that, the expressions in Eqs. (\ref{eq:me_11_S_1})
and (\ref{eq:me_11_S_2}) are equivalent. In the actual calculation
below, we will adopt the latter expression as it has a clearer meaning
and closely corresponds to the calculation under the diquark picture,
see, for example, Ref. \cite{Wang:2017mqp}.

Some comments are in order. In Ref. \cite{Wang:2017mqp}, the so-called
``overlap factor'' is introduced, which is actually the inner product
of the flavor-spin wavefunctions of the initial and final states. However,
in the diquark picture, the unclear definition of diquark makes it
difficult to determine this factor. It is obvious that this factor
is very important because it appears directly as a multiplication
factor in the calculation of matrix elements. This factor is also
present in the three-quark picture. For example, in Eq. (\ref{eq:me_11_S_1}),
this factor is $(-1)\times1/\sqrt{2}\times2$, while in Eq. (\ref{eq:me_11_S_2}),
it is $(-\sqrt{3}/2)\times1/\sqrt{2}\times2$. It can be seen that
if different calculation schemes are used, this factor can also be
different. Readers who want to use this method must be careful enough.
More discussion can be found in Subsec. \ref{subsec:overlap_factor}.

The form factors $f_{i}$ can be extracted in the following way. Respectively
multiply Eq.~(\ref{eq:parameterization_11_V}) by $\sum_{S_{z},S_{z}^{\prime}}\bar{u}(P,S_{z})\Gamma_{\mu}u(P^{\prime},S_{z}^{\prime})$
with $\Gamma_{\mu}\in\{\gamma_{\mu},P_{\mu},P_{\mu}^{\prime}\}$ to
arrive at one set of expressions. Do the same thing for the vector-current
part of Eq.~(\ref{eq:me_11_S_2}), but at this point, take
the approximation $P^{(\prime)}\to\bar{P}^{(\prime)}$ within the
integral. Equate the two sets of expressions to extract the form factors
$f_{i}$. Similarly, one can also extract the form factors $g_{i}$.

\subsection{The $1/2\to1/2$ case with an axial-vector spectator diquark}

In this subsection, we will take $\Omega_{bcc}^{+}\to\Omega_{bc}^{\prime0}$
as an example to illustrate how to extract the form factors of the
$1/2\to1/2$ case with an axial-vector spectator diquark. This is
also a $c\to s$ process, and the $b$ quark and one $c$ quark are
spectators. For the initial state, the two charm quarks are considered
to form an axial-vector diquark, while for the final state, the two
heavy quarks are considered to form an axial-vector diquark. 

One the one hand, the weak decay matrix element can be parameterized
in terms of form factors, see Eqs. (\ref{eq:parameterization_11_V})
and (\ref{eq:parameterization_11_A}). On the other hand, the matrix
element can also be calculated in LFQM in two different ways.

\paragraph*{(1) The direct way}

Doing the contraction directly, one can obtain
\begin{align}
 & \langle\Omega_{bc}^{\prime0}(P^{\prime},S_{z}^{\prime})|\bar{s}\gamma^{\mu}(1-\gamma_{5})c|\Omega_{bcc}^{+}(P,S_{z})\rangle\nonumber \\
= & \int\{d^{3}\tilde{p}_{2}\}\{d^{3}\tilde{p}_{3}\}\frac{A^{\prime}A}{\sqrt{p_{1}^{\prime+}p_{1}^{+}P^{\prime+}P^{+}}}\Phi^{\prime*}(x_{i}^{\prime},k_{i\perp}^{\prime})\Phi(x_{i},k_{i\perp})\times\frac{1}{\sqrt{2}}\times2\nonumber \\
 & \times\bar{u}(\bar{P}^{\prime},S_{z}^{\prime})(\frac{1}{\sqrt{3}}\gamma_{\sigma}\gamma_{5})(\slashed p_{1}^{\prime}+m_{1}^{\prime})\gamma^{\mu}(1-\gamma_{5})(\slashed p_{1}+m_{1})(\bar{\slashed P}+M_{0})(\gamma^{\rho}-v^{\rho})\nonumber \\
 & \times C(\slashed p_{3}+m_{3})^{T}C(\gamma^{\sigma}-v^{\prime\sigma})(\bar{\slashed P^{\prime}}+M_{0}^{\prime})(\slashed p_{2}+m_{2})(\frac{1}{\sqrt{3}}\gamma_{\rho}\gamma_{5})u(\bar{P},S_{z}),\label{eq:me_11_A_1}
\end{align}
where the four quarks can also be numbered as $(1,1^{\prime},2,3)=(c^{1},s,b,c^{2})$
with $c^{i}$ the $i$-th charm quark.

\paragraph*{(2) The indirect way}

The spin wavefunction of the initial state has been expanded in terms
of a new diquark basis in Eq. (\ref{eq:ccb_2_cbc}), while the spin
wavefunction of the final state is $\psi_{1}(cbs)$, then one can
read the factor $-1/2$. It turns out that
\begin{align}
 & \langle\Omega_{bc}^{\prime0}(P^{\prime},S_{z}^{\prime})|\bar{s}\gamma^{\mu}(1-\gamma_{5})c|\Omega_{bcc}^{+}(P,S_{z})\rangle\nonumber \\
= & \int\{d^{3}\tilde{p}_{2}\}\{d^{3}\tilde{p}_{3}\}\frac{A^{\prime}A}{\sqrt{p_{1}^{\prime+}p_{1}^{+}P^{\prime+}P^{+}}}\Phi^{\prime*}(x_{i}^{\prime},k_{i\perp}^{\prime})\Phi(x_{i},k_{i\perp})\times(-\frac{1}{2})\times\frac{1}{\sqrt{2}}\times2\nonumber \\
 & \times{\rm Tr}[(\bar{\slashed P}+M_{0})(\gamma^{\rho}-v^{\rho})C(\slashed p_{2}+m_{2})^{T}C(\gamma^{\sigma}-v^{\prime\sigma})(\bar{\slashed P^{\prime}}+M_{0}^{\prime})(\slashed p_{3}+m_{3})]\nonumber \\
 & \times\bar{u}(\bar{P}^{\prime},S_{z}^{\prime})(\frac{1}{\sqrt{3}}\gamma_{\sigma}\gamma_{5})(\slashed p_{1}^{\prime}+m_{1}^{\prime})\gamma^{\mu}(1-\gamma_{5})(\slashed p_{1}+m_{1})(\frac{1}{\sqrt{3}}\gamma_{\rho}\gamma_{5})u(\bar{P},S_{z}).\label{eq:me_11_A_2}
\end{align}

One can check that, the expressions in Eqs. (\ref{eq:me_11_A_1})
and (\ref{eq:me_11_A_2}) are equivalent. In the actual calculation
below, we will adopt the latter expression.

The method for extracting the form factors is the same as that in
Subsec. \ref{subsec:ff_11_S}.

\subsection{The overlap factor}

\label{subsec:overlap_factor}

As mentioned above, the overlap factor is so important that it is
worth further discussing in a new subsection. It must be emphasized again
that the overlap factor may be different if different calculation
schemes are used. For the form factors of the $1/2\to1/2$ case, regardless
of whether the spectator diquark is a scalar or an axial-vector, we
all adopt the indirect calculation schemes. 

In the three-quark picture, the overlap factor consists of the following
parts:
\begin{itemize}
\item inner product in spin space,
\item the normalization factor of initial state,
\item the normalization factor of final state,
\item contraction factor.
\end{itemize}
Of course, the latter three items originate from the symmetry in flavor
space. The overlap factors of the three typical processes have been
given above. Here, we only present another example, where the initial state
contains more identical quarks -- $\Omega_{ccc}^{++}(ccc)\to\Omega_{cc}^{+}(scc)$:
\begin{itemize}
\item For the initial state, whether in flavor space or spin space, the
three charm quarks are on an equal footing. Especially, one can consider
the two spectator charm quarks as an axial-vector diquark. For the
final state, the two spectator charm quarks form an axial-vector diquark.
Therefore, the inner product in spin space is 1.
\item The normalization factor of initial state is $1/\sqrt{6}$. This is
because when calculating $\langle\Omega_{ccc}^{++}|\Omega_{ccc}^{++}\rangle$,
a contraction factor 6 appears, therefore, in order to normalize the flavor wavefunction
of $|\Omega_{ccc}^{++}\rangle$, an additional factor $1/\sqrt{6}$
should be added.
\item The normalization factor of final state is $1/\sqrt{2}$. Similar
to the last item.
\item The contraction factor of $\langle\Omega_{cc}^{+}|\bar{s}\gamma^{\mu}(1-\gamma_{5})c|\Omega_{ccc}^{++}\rangle$
is 6. 
\end{itemize}
The overlap factors of other processes can be found in Table \ref{Tab:overlap_factors}.

\begin{table}
\caption{Overlap factors in flavor-spin space.}
\label{Tab:overlap_factors}
\begin{tabular}{c|c|c|c}
\hline 
$3/2\to1/2$ transition & Overlap factor & $1/2\to1/2$ transition & Overlap factor\tabularnewline
\hline 
$\Omega_{ccc}^{++}(ccc)\to\Xi_{cc}^{+}(dcc)/\Omega_{cc}^{+}(scc)$ & $1\times\frac{1}{\sqrt{6}}\times\frac{1}{\sqrt{2}}\times6$ & $\Omega_{bcc}^{+}(ccb)\to\Xi_{bc}^{0}(dcb)/\Omega_{bc}^{0}(scb)$ & $(-\frac{\sqrt{3}}{2})\times\frac{1}{\sqrt{2}}\times1\times2$\tabularnewline
$\Omega_{bcc}^{\prime+}(ccb)\to\Xi_{bc}^{\prime0}(dcb)/\Omega_{bc}^{\prime0}(scb)$ & $1\times\frac{1}{\sqrt{2}}\times1\times2$ & $\Omega_{bbc}^{0}(bbc)\to\Xi_{bc}^{+}(ubc)$ & $(-\frac{\sqrt{3}}{2})\times\frac{1}{\sqrt{2}}\times1\times2$\tabularnewline
\cline{3-4} \cline{4-4} 
$\Omega_{bbc}^{\prime0}(cbb)\to\Xi_{bb}^{-}(dbb)/\Omega_{bb}^{-}(sbb)$ & $1\times\frac{1}{\sqrt{2}}\times\frac{1}{\sqrt{2}}\times2$ & $\Omega_{bcc}^{+}(ccb)\to\Xi_{bc}^{\prime0}(dcb)/\Omega_{bc}^{\prime0}(scb)$ & $(-\frac{1}{2})\times\frac{1}{\sqrt{2}}\times1\times2$\tabularnewline
$\Omega_{bcc}^{\prime+}(bcc)\to\Xi_{cc}^{++}(ucc)$ & $1\times\frac{1}{\sqrt{2}}\times\frac{1}{\sqrt{2}}\times2$ & $\Omega_{bbc}^{0}(cbb)\to\Xi_{bb}^{-}(dbb)/\Omega_{bb}^{-}(sbb)$ & $1\times\frac{1}{\sqrt{2}}\times\frac{1}{\sqrt{2}}\times2$\tabularnewline
$\Omega_{bbc}^{\prime0}(bbc)\to\Xi_{bc}^{\prime+}(ubc)$ & $1\times\frac{1}{\sqrt{2}}\times1\times2$ & $\Omega_{bcc}^{+}(bcc)\to\Xi_{cc}^{++}(ucc)$ & $1\times\frac{1}{\sqrt{2}}\times\frac{1}{\sqrt{2}}\times2$\tabularnewline
$\Omega_{bbb}^{-}(bbb)\to\Xi_{bb}^{0}(ubb)$ & $1\times\frac{1}{\sqrt{6}}\times\frac{1}{\sqrt{2}}\times6$ & $\Omega_{bbc}^{0}(bbc)\to\Xi_{bc}^{\prime+}(ubc)$ & $(-\frac{1}{2})\times\frac{1}{\sqrt{2}}\times1\times2$\tabularnewline
\hline 
\end{tabular}
\end{table}

\section{Numerical results and phenomenological applications}

In this section, we will first present the numerical results of the
form factors. Subsequently, these form factors will be applied to arrive at
some phenomenological predictions, including semileptonic decays and
nonleptonic decays. For the latter, we are constrained to consider
only the factorable W-emission diagram, i.e., the diagram with a W
boson emitting outward and decaying into a meson. It should be noted that, nonleptonic
decays contain additional non-perturbative contributions, our predictions
can only be considered as rough estimates. However, considering
nonleptonic decays have practical significance for the experimental
search for THBs, we still believe that our estimates here are valuable.

\subsection{Inputs}

The masses of the initial THBs and the final DHBs are collected in
Table \ref{Tab:masses_poles}. The former come from the QCD sum rules
calculation \cite{Wang:2020avt}, while the latter come from the experimental
measurement \cite{LHCb:2017iph} and the lattice QCD calculation \cite{Brown:2014ena}.
In addition, we will take $m_{\Xi_{bc}}=m_{\Xi_{bc}^{\prime}}$and
$m_{\Omega_{bc}}=m_{\Omega_{bc}^{\prime}}$.

In Table \ref{Tab:masses_poles}, we also list the pole residues of
THBs and DHBs, which are respectively calculated in Refs. \cite{Wang:2020avt}
and \cite{Hu:2017dzi} in QCD sum rules. With the help of pole residues,
we determine the shape parameters, which are also listed in Table
\ref{Tab:masses_poles}. One can see from Table \ref{Tab:masses_poles}
that, although the pole residues of spin-3/2 $\Omega_{bcc}^{\prime+}$
($\Omega_{bbc}^{\prime0}$) and spin-1/2 $\Omega_{bcc}^{+}$ ($\Omega_{bbc}^{0}$)
differ greatly, the shape parameters determined by them are almost
equal, which is as expected. In this sense, taking $\beta_{\Xi_{bc}}=\beta_{\Xi_{bc}^{\prime}}$
and $\beta_{\Omega_{bc}}=\beta_{\Omega_{bc}^{\prime}}$ is reasonable.

\begin{table}
\caption{Masses and pole residues of THBs and DHBs, as well as shape parameters
determined by pole residues.}
\label{Tab:masses_poles} %
\begin{tabular}{l|c|c|c|c|c|c}
\hline 
 & $\Omega_{ccc}^{++}$ & $\Omega_{bcc}^{\prime+}$ & $\Omega_{bbc}^{\prime0}$ & $\Omega_{bbb}^{-}$ & $\Omega_{bcc}^{+}$ & $\Omega_{bbc}^{0}$\tabularnewline
\hline 
$m/{\rm GeV}$ & $4.81$ & $8.03$ & $11.23$ & $14.43$ & $8.02$ & $11.22$\tabularnewline
$\lambda/{\rm GeV}^{3}$ & $0.208\pm0.031$ & $0.225\pm0.025$ & $0.324\pm0.046$ & $0.942\pm0.139$ & $0.430\pm0.047$ & $0.565\pm0.081$\tabularnewline
$\beta/{\rm GeV}$ & $0.704\pm0.036$ & $0.869\pm0.032$ & $0.970\pm0.045$ & $1.134\pm0.054$ & $0.890\pm0.033$ & $0.966\pm0.046$\tabularnewline
\hline 
 & $\Xi_{cc}$ & $\Omega_{cc}$ & $\Xi_{bc}^{\prime}$ & $\Omega_{bc}^{\prime}$ & $\Xi_{bb}$ & $\Omega_{bb}$\tabularnewline
\hline 
$m/{\rm GeV}$ & $3.622$ & $3.738$ & $6.943$ & $6.998$ & $10.143$ & $10.273$\tabularnewline
$\lambda/{\rm GeV}^{3}$ & $0.109\pm0.021$ & $0.123\pm0.024$ & $0.176\pm0.040$ & $0.188\pm0.041$ & $0.281\pm0.071$ & $0.347\pm0.083$\tabularnewline
$\beta/{\rm GeV}$ & $0.583\pm0.037$ & $0.600\pm0.039$ & $0.687\pm0.051$ & $0.694\pm0.050$ & $0.803\pm0.065$ & $0.853\pm0.065$\tabularnewline
\hline 
\end{tabular}
\end{table}

The constituent quark masses are adopted as (in units of GeV)~\cite{Lu:2007sg,Wang:2007sxa,Wang:2008xt,Wang:2008ci,Wang:2009mi,Chen:2009qk,Li:2010bb,Verma:2011yw,Shi:2016gqt}
\begin{equation}
m_{u}=m_{d}=0.25,\quad m_{s}=0.37,\quad m_{c}=1.4,\quad m_{b}=4.8.
\end{equation}

The decay constants of mesons in the final states of nonleptonic decays
can be found in Refs.~\cite{Zhao:2018mrg,Cheng:2003sm,Shi:2016gqt,Carrasco:2014poa}.
The Wilson coefficient $a_{1}(\mu)\equiv C_{1}(\mu)+C_{2}(\mu)/3$
is taken as $a_{1}(\mu_{b})=1.03$ for the bottom quark decay and
$a_{1}(\mu_{c})=1.10$ for the charmed quark decay~\cite{Buras:1998raa}.
All the other inputs can be found in PDG~\cite{ParticleDataGroup:2022pth}.

\subsection{Form factors}

Our predictions for the form factors are given in Tables \ref{Tab:ff31},
\ref{Tab:ff11S}, and \ref{Tab:ff11A}, where the values of the form
factors at $q^{2}=0$ are shown. At this point, the SU(3) flavor symmetry and heavy
quark symmetry, as well as their breaking degrees can be seen most clearly.
Some comments are in order.
\begin{itemize}
\item The $c\to d$ and $c\to s$ processes
are related by SU(3) flavor symmetry. One can see that the SU(3) symmetry
breaking is roughly 10-20\%. 
\item For example, $\Omega_{ccc}^{++}\to\Xi_{cc}^{+}$ and $\Omega_{bbb}^{-}\to\Xi_{bb}^{0}$
are related by heavy quark symmetry. One can see that this symmetry
is bad, which should be mainly attributed to the fact that the charm
quark is not heavy enough. 
\item Comparing Table \ref{Tab:ff11S} and Table \ref{Tab:ff11A}, we find
that the form factors involving a scalar spectator diquark are larger than the corresponding ones
involving an axial-vector spectator diquark.
\end{itemize}
To access the $q^{2}$ dependence in the kinematic region, we adopt
the following single pole assumption:
\begin{equation}
F(q^{2})=\frac{F(0)}{1-q^{2}/m_{{\rm pole}}^{2}}.\label{eq:F_q2}
\end{equation}
For the $c\to d/s$ and $b\to u$ decays, $m_{{\rm pole}}$ is respectively
taken as the mass of $D$, $D_{s}$, and $B$ meson. For a discussion
on the validity of this assumption, see Ref. \cite{Shi:2016gqt}.

\begin{table}
\caption{Form factors at $q^{2}=0$ for the $3/2\to1/2$ processes. }
\label{Tab:ff31} %
\begin{tabular}{l|r|r|r|r|r|r|r|r}
\hline 
Transition & $f_{1}^{V}(0)$ & $f_{2}^{V}(0)$ & $f_{3}^{V}(0)$ & $f_{4}^{V}(0)$ & $f_{1}^{A}(0)$ & $f_{2}^{A}(0)$ & $f_{3}^{A}(0)$ & $f_{4}^{A}(0)$\tabularnewline
\hline 
$\Omega_{ccc}^{++}\to\Xi_{cc}^{+}$ & $-1.411$ & $-1.275$ & $-0.119$ & $2.879$ & $7.478$ & $-13.167$ & $5.070$ & $1.263$\tabularnewline
$\Omega_{bcc}^{\prime+}\to\Xi_{bc}^{\prime0}$ & $-2.314$ & $-0.419$ & $-1.851$ & $4.621$ & $18.746$ & $-32.808$ & $13.621$ & $1.009$\tabularnewline
$\Omega_{bbc}^{\prime0}\to\Xi_{bb}^{-}$ & $-2.471$ & $2.610$ & $-5.039$ & $4.907$ & $35.792$ & $-57.419$ & $21.567$ & $0.827$\tabularnewline
\hline 
$\Omega_{ccc}^{++}\to\Omega_{cc}^{+}$ & $-1.643$ & $-1.653$ & $0.025$ & $3.349$ & $8.314$ & $-14.675$ & $5.866$ & $1.363$\tabularnewline
$\Omega_{bcc}^{\prime+}\to\Omega_{bc}^{\prime0}$ & $-2.431$ & $-0.315$ & $-2.072$ & $4.854$ & $23.062$ & $-37.360$ & $14.200$ & $1.123$\tabularnewline
$\Omega_{bbc}^{\prime0}\to\Omega_{bb}^{-}$ & $-2.954$ & $3.587$ & $-6.492$ & $5.872$ & $37.389$ & $-55.349$ & $18.426$ & $0.863$\tabularnewline
\hline 
$\Omega_{bcc}^{\prime+}\to\Xi_{cc}^{++}$ & $-0.033$ & $-0.027$ & $-0.004$ & $0.073$ & $0.026$ & $-0.076$ & $0.018$ & $0.047$\tabularnewline
$\Omega_{bbc}^{\prime0}\to\Xi_{bc}^{\prime+}$ & $-0.037$ & $-0.023$ & $-0.012$ & $0.074$ & $0.060$ & $-0.151$ & $0.055$ & $0.034$\tabularnewline
$\Omega_{bbb}^{-}\to\Xi_{bb}^{0}$ & $-0.123$ & $-0.051$ & $-0.067$ & $0.244$ & $0.275$ & $-0.684$ & $0.298$ & $0.086$\tabularnewline
\hline 
\end{tabular}
\end{table}

\begin{table}
\caption{Form factors for the $1/2\to1/2$ processes with a scalar spectator diquark.}
\label{Tab:ff11S}%
\begin{tabular}{l|r|r|r|r|r|r}
\hline 
Transition & $f_{1}(0)$ & $f_{2}(0)$ & $f_{3}(0)$ & $g_{1}(0)$ & $g_{2}(0)$ & $g_{3}(0)$\tabularnewline
\hline 
$\Omega_{bcc}^{+}\to\Xi_{bc}^{0}$ & $-0.582$ & $2.932$ & $-1.106$ & $-0.396$ & $-2.141$ & $29.448$\tabularnewline
\hline 
$\Omega_{bcc}^{+}\to\Omega_{bc}^{0}$ & $-0.630$ & $3.269$ & $-1.500$ & $-0.445$ & $-2.479$ & $31.279$\tabularnewline
\hline 
$\Omega_{bbc}^{0}\to\Xi_{bc}^{+}$ & $-0.021$ & $0.026$ & $0.003$ & $-0.015$ & $-0.022$ & $0.179$\tabularnewline
\hline 
\end{tabular}
\end{table}

\begin{table}
\caption{Form factors for the $1/2\to1/2$ processes with an axial-vector spectator diquark.}
\label{Tab:ff11A}%
\begin{tabular}{l|r|r|r|r|r|r}
\hline 
Transition & $f_{1}(0)$ & $f_{2}(0)$ & $f_{3}(0)$ & $g_{1}(0)$ & $g_{2}(0)$ & $g_{3}(0)$\tabularnewline
\hline 
$\Omega_{bcc}^{+}\to\Xi_{bc}^{\prime0}$ & $-0.335$ & $0.237$ & $-0.539$ & $0.076$ & $0.411$ & $-5.648$\tabularnewline
$\Omega_{bbc}^{0}\to\Xi_{bb}^{-}$ & $0.526$ & $-1.804$ & $1.649$ & $-0.112$ & $-1.054$ & $19.543$\tabularnewline
\hline 
$\Omega_{bcc}^{+}\to\Omega_{bc}^{\prime0}$ & $-0.363$ & $0.361$ & $-0.769$ & $0.086$ & $0.476$ & $-6.001$\tabularnewline
$\Omega_{bbc}^{0}\to\Omega_{bb}^{-}$ & $0.610$ & $-2.710$ & $1.947$ & $-0.143$ & $-1.010$ & $17.581$\tabularnewline
\hline 
$\Omega_{bcc}^{+}\to\Xi_{cc}^{++}$ & $0.035$ & $0.036$ & $-0.023$ & $-0.009$ & $-0.007$ & $0.036$\tabularnewline
$\Omega_{bbc}^{0}\to\Xi_{bc}^{\prime+}$ & $-0.012$ & $-0.010$ & $0.007$ & $0.003$ & $0.004$ & $-0.034$\tabularnewline
\hline 
\end{tabular}
\end{table}

\subsection{Semileptonic decays}

\subsubsection{The $3/2^{+}\to1/2^{+}$ case}

Define the helicity amplitudes as
\begin{equation}
H_{\lambda^{\prime},\lambda_{W}}^{V(A)}\equiv\langle{\cal B}^{\prime}(\lambda^{\prime})|\bar{q}\gamma^{\mu}(\gamma_{5})Q|{\cal B}(\lambda)\rangle\epsilon_{W\mu}^{*}(\lambda_{W}),
\label{eq:HA}
\end{equation}
where $\lambda=\lambda_{W}-\lambda^{\prime}$. These amplitudes can
be written in terms of form factors: 
\begin{align}
H_{\frac{1}{2},1}^{V,A} & =\mp i\sqrt{\frac{1}{3}}\sqrt{2MM^{\prime}(\omega\mp1)}\Big[2(\pm\omega+1)f_{1}^{V,A}+f_{4}^{V,A}\Big],\nonumber \\
H_{\frac{1}{2},-1}^{V,A} & =\mp i\sqrt{2MM^{\prime}(\omega\mp1)}f_{4}^{V,A},\nonumber \\
H_{\frac{1}{2},0}^{V,A} & =\pm i\sqrt{\frac{2}{3}}\frac{\sqrt{2MM^{\prime}(\omega\mp1)}}{\hat{q}}\Big[(\mp1+r)(\omega\pm1)f_{1}^{V,A}+r(\omega^{2}-1)f_{2}^{V,A}\nonumber \\
 & \qquad\qquad\qquad\qquad\qquad+(\omega^{2}-1)f_{3}^{V,A}+(r\omega-1)f_{4}^{V,A}\Big],\nonumber \\
H_{\frac{1}{2},t}^{V,A} & =\mp i\sqrt{\frac{2}{3}}(\omega\mp1)\frac{\sqrt{2MM^{\prime}(\omega\pm1)}}{\hat{q}}\Big[(\pm1+r)f_{1}^{V,A}+(r\omega-1)f_{2}^{V,A}\nonumber \\
 & \qquad\qquad\qquad\qquad\qquad+(r-\omega)f_{3}^{V,A}+rf_{4}^{V,A}\Big],
\end{align}
where the upper (lower) sign corresponds to $V$ ($A$), $\omega\equiv v\cdot v^{\prime}=(P\cdot P^{\prime})/(MM^{\prime})$,
$r\equiv M^{\prime}/M$, and $\hat{q}\equiv\sqrt{q^{2}}/M$. The other
helicity amplitudes can be obtained by
\begin{equation}
H_{-\lambda^{\prime},-\lambda_{W}}^{V,A}=\mp H_{\lambda^{\prime},\lambda_{W}}^{V,A}.
\end{equation}

The polarized differential decay widths for ${\cal B}(3/2^{+})\to{\cal B}^{\prime}(1/2^{+})l\nu$
are
\begin{align}
\frac{d\Gamma_{L}}{dq^{2}} & =\frac{G_{F}^{2}|V_{{\rm CKM}}|^{2}|\vec{P}^{\prime}|q^{2}(1-\hat{m}_{l}^{2})^{2}}{768\pi^{3}M^{2}}\left[(2+\hat{m}_{l}^{2})(|H_{-\frac{1}{2},0}|^{2}+|H_{\frac{1}{2},0}|^{2})+3\hat{m}_{l}^{2}(|H_{-\frac{1}{2},t}|^{2}+|H_{\frac{1}{2},t}|^{2})\right],\nonumber \\
\frac{d\Gamma_{T}}{dq^{2}} & =\frac{G_{F}^{2}|V_{{\rm CKM}}|^{2}|\vec{P}^{\prime}|q^{2}(1-\hat{m}_{l}^{2})^{2}(2+\hat{m}_{l}^{2})}{768\pi^{3}M^{2}}\left[|H_{-\frac{1}{2},-1}|^{2}+|H_{-\frac{1}{2},1}|^{2}+|H_{\frac{1}{2},-1}|^{2}+|H_{\frac{1}{2},1}|^{2}\right],
\end{align}
where $H_{\lambda^{\prime},\lambda_{W}}\equiv H_{\lambda^{\prime},\lambda_{W}}^{V}-H_{\lambda^{\prime},\lambda_{W}}^{A}$,
$\hat{m}_{l}\equiv m_{l}/\sqrt{q^{2}}$, and $|\vec{P}^{\prime}|$
is the magnitude of 3-momentum of ${\cal B}^{\prime}$ in the rest
frame of ${\cal B}$.

Our predictions for the semileptonic decays of the $3/2\to1/2$ processes
are given in Table \ref{Tab:semi_31}. One can see that, the decay
widths of the $c\to d$ processes are almost exactly one order
of magnitude smaller than the corresponding ones of the $c\to s$ processes,
due to $|V_{cd}/V_{cs}|^{2}\approx0.054$. After considering these
CKM matrix elements, we find that, the SU(3) symmetry breaking between
$\Omega_{ccc}^{++}\to\Xi_{cc}^{+}e^{+}\nu_{e}$ and $\Omega_{ccc}^{++}\to\Omega_{cc}^{+}e^{+}\nu_{e}$
is about 30\%, that between $\Omega_{bcc}^{\prime+}\to\Xi_{bc}^{\prime0}e^{+}\nu_{e}$
and $\Omega_{bcc}^{\prime+}\to\Omega_{bc}^{\prime0}e^{+}\nu_{e}$
is about 10\%, and that between $\Omega_{bbc}^{\prime0}\to\Xi_{bb}^{-}e^{+}\nu_{e}$
and $\Omega_{bbc}^{\prime0}\to\Omega_{bb}^{-}e^{+}\nu_{e}$ is about
40\%. 

\begin{table}
\caption{Decay widths and $\Gamma_{L}/\Gamma_{T}$ for the semileptonic decays
of the $3/2\to1/2$ processes.}
\label{Tab:semi_31} %
\begin{tabular}{l|c|c}
\hline 
Channel & $\Gamma/\text{~GeV}$ & $\Gamma_{L}/\Gamma_{T}$\tabularnewline
\hline 
$\Omega_{ccc}^{++}\to\Xi_{cc}^{+}e^{+}\nu_{e}$ & $1.11\times10^{-14}$ & $0.58$\tabularnewline
$\Omega_{bcc}^{\prime+}\to\Xi_{bc}^{\prime0}e^{+}\nu_{e}$ & $5.18\times10^{-15}$ & $0.63$\tabularnewline
$\Omega_{bbc}^{\prime0}\to\Xi_{bb}^{-}e^{+}\nu_{e}$ & $3.59\times10^{-15}$ & $0.60$\tabularnewline
\hline 
$\Omega_{ccc}^{++}\to\Omega_{cc}^{+}e^{+}\nu_{e}$ & $1.31\times10^{-13}$ & $0.64$\tabularnewline
$\Omega_{bcc}^{\prime+}\to\Omega_{bc}^{\prime0}e^{+}\nu_{e}$ & $8.44\times10^{-14}$ & $0.65$\tabularnewline
$\Omega_{bbc}^{\prime0}\to\Omega_{bb}^{-}e^{+}\nu_{e}$ & $3.60\times10^{-14}$ & $0.69$\tabularnewline
\hline 
$\Omega_{bcc}^{\prime+}\to\Xi_{cc}^{++}e^{-}\bar{\nu}_{e}$ & $3.15\times10^{-18}$ & $0.65$\tabularnewline
$\Omega_{bbc}^{\prime0}\to\Xi_{bc}^{\prime+}e^{-}\bar{\nu}_{e}$ & $1.89\times10^{-18}$ & $0.63$\tabularnewline
$\Omega_{bbb}^{-}\to\Xi_{bb}^{0}e^{-}\bar{\nu}_{e}$ & $1.41\times10^{-17}$ & $0.62$\tabularnewline
\hline 
\end{tabular}
\end{table}

\subsubsection{The $1/2^{+}\to1/2^{+}$ case}

Similarly define the helicity amplitudes as in Eq. (\ref{eq:HA}). These amplitudes
can also be written in terms of form factors: 
\begin{align}
 & H_{\frac{1}{2},1}^{V}=-i\sqrt{2Q_{-}}\left(f_{1}-\frac{M+M^{\prime}}{M}f_{2}\right),\quad H_{\frac{1}{2},0}^{V}=-i\frac{\sqrt{Q_{-}}}{\sqrt{q^{2}}}\left((M+M^{\prime})f_{1}-\frac{q^{2}}{M}f_{2}\right),\nonumber \\
 & H_{\frac{1}{2},t}^{V}=-i\frac{\sqrt{Q_{+}}}{\sqrt{q^{2}}}\left((M-M^{\prime})f_{1}+\frac{q^{2}}{M}f_{3}\right),\nonumber \\
 & H_{\frac{1}{2},1}^{A}=-i\sqrt{2Q_{+}}\left(g_{1}+\frac{M-M^{\prime}}{M}g_{2}\right),\quad H_{\frac{1}{2},0}^{A}=-i\frac{\sqrt{Q_{+}}}{\sqrt{q^{2}}}\left((M-M^{\prime})g_{1}+\frac{q^{2}}{M}g_{2}\right),\nonumber \\
 & H_{\frac{1}{2},t}^{A}=-i\frac{\sqrt{Q_{-}}}{\sqrt{q^{2}}}\left((M+M^{\prime})g_{1}-\frac{q^{2}}{M}g_{3}\right),
\end{align}
where $Q_{\pm}=(M\pm M^{\prime})^{2}-q^{2}$ and $M^{(\prime)}$ is
the mass of the initial (final) baryon. The other helicity amplitudes
can be obtained by
\begin{equation}
H_{-\lambda^{\prime},-\lambda_{W}}^{V,A}=\pm H_{\lambda^{\prime},\lambda_{W}}^{V,A}.
\end{equation}

The polarized differential decay widths for ${\cal B}(1/2^{+})\to{\cal B}^{\prime}(1/2^{+})l\nu$
are
\begin{align}
\frac{d\Gamma_{L}}{dq^{2}} & =\frac{G_{F}^{2}|V_{{\rm CKM}}|^{2}|\vec{P}^{\prime}|q^{2}(1-\hat{m}_{l}^{2})^{2}}{384\pi^{3}M^{2}}\left[(2+\hat{m}_{l}^{2})(|H_{-\frac{1}{2},0}|^{2}+|H_{\frac{1}{2},0}|^{2})+3\hat{m}_{l}^{2}(|H_{-\frac{1}{2},t}|^{2}+|H_{\frac{1}{2},t}|^{2})\right],\nonumber \\
\frac{d\Gamma_{T}}{dq^{2}} & =\frac{G_{F}^{2}|V_{{\rm CKM}}|^{2}|\vec{P}^{\prime}|q^{2}(1-\hat{m}_{l}^{2})^{2}(2+\hat{m}_{l}^{2})}{384\pi^{3}M^{2}}\left[|H_{-\frac{1}{2},-1}|^{2}+|H_{\frac{1}{2},1}|^{2}\right],
\end{align}
where $H_{\lambda^{\prime},\lambda_{W}}\equiv H_{\lambda^{\prime},\lambda_{W}}^{V}-H_{\lambda^{\prime},\lambda_{W}}^{A}$,
$\hat{m}_{l}\equiv m_{l}/\sqrt{q^{2}}$, and $|\vec{P}^{\prime}|$
is the magnitude of 3-momentum of ${\cal B}^{\prime}$ in the rest
frame of ${\cal B}$.

Our predictions for the semileptonic decays of the $1/2\to1/2$ processes
are given in Tables \ref{Tab:semi_11S} and \ref{Tab:semi_11A}. Some
comments are in order.
\begin{itemize}
\item As expected, the decay widths of the $c\to d$ processes are almost
exactly one order of magnitude smaller than the corresponding ones of the $c\to s$
processes. 
\item The decay widths involving a scalar spectator diquark are almost exactly
one order of magnitude larger than the corresponding ones involving an axial-vector spectator
diquark. 
\item Our predictions on $\Gamma_{L}/\Gamma_{T}$ are quite interesting.
For the decay widths involving a scalar spectator diquark, $\Gamma_{L}\approx\Gamma_{T}$,
while for those involving an axial-vector spectator diquark, $\Gamma_{L}\gg\Gamma_{T}$.
Specifically, take the two decays $\Omega_{bcc}^{+}\to\Xi_{bc}^{(\prime)0}e^{+}\nu_{e}$
as an example. Both of these two decays are the $1/2^{+}\to1/2^{+}$
semileptonic decays, and the only difference between them is that
the two final states have different flavor-spin wavefunctions. It seems that $\Gamma_{L}/\Gamma_{T}$
can tell us some information about the internal structures of $\Xi_{bc}^{(\prime)0}$. 
\end{itemize}

\begin{table}
\caption{Decay widths and $\Gamma_{L}/\Gamma_{T}$ for the semileptonic decays
of the $1/2\to1/2$ processes with a scalar spectator diquark.}
\label{Tab:semi_11S} %
\begin{tabular}{l|c|c}
\hline 
Channel & $\Gamma/\text{~GeV}$ & $\Gamma_{L}/\Gamma_{T}$\tabularnewline
\hline 
$\Omega_{bcc}^{+}\to\Xi_{bc}^{0}e^{+}\nu_{e}$ & $9.67\times10^{-15}$ & $0.96$\tabularnewline
\hline 
$\Omega_{bcc}^{+}\to\Omega_{bc}^{0}e^{+}\nu_{e}$ & $1.57\times10^{-13}$ & $0.98$\tabularnewline
\hline 
$\Omega_{bbc}^{0}\to\Xi_{bc}^{+}e^{-}\bar{\nu}_{e}$ & $3.34\times10^{-18}$ & $0.89$\tabularnewline
\hline 
\end{tabular}
\end{table}

\begin{table}
\caption{Decay widths and $\Gamma_{L}/\Gamma_{T}$ for the semileptonic decays
of the $1/2\to1/2$ processes with an axial-vector spectator diquark.}
\label{Tab:semi_11A} %
\begin{tabular}{l|c|c}
\hline 
Channel & $\Gamma/\text{~GeV}$ & $\Gamma_{L}/\Gamma_{T}$\tabularnewline
\hline 
$\Omega_{bcc}^{+}\to\Xi_{bc}^{\prime0}e^{+}\nu_{e}$ & $8.91\times10^{-16}$ & $4.29$\tabularnewline
$\Omega_{bbc}^{0}\to\Xi_{bb}^{-}e^{+}\nu_{e}$ & $2.49\times10^{-15}$ & $3.22$\tabularnewline
\hline 
$\Omega_{bcc}^{+}\to\Omega_{bc}^{\prime0}e^{+}\nu_{e}$ & $1.45\times10^{-14}$ & $4.19$\tabularnewline
$\Omega_{bbc}^{0}\to\Omega_{bb}^{-}e^{+}\nu_{e}$ & $2.98\times10^{-14}$ & $3.59$\tabularnewline
\hline 
$\Omega_{bcc}^{+}\to\Xi_{cc}^{++}e^{-}\bar{\nu}_{e}$ & $1.83\times10^{-18}$ & $3.15$\tabularnewline
$\Omega_{bbc}^{0}\to\Xi_{bc}^{\prime+}e^{-}\bar{\nu}_{e}$ & $2.57\times10^{-19}$ & $3.38$\tabularnewline
\hline 
\end{tabular}
\end{table}

\subsection{Nonleptonic decays}

In this subsection, we will only consider the contribution from the
factorable W-emission diagram, in which the W boson is emitted outward
and decays into a pseudoscalar or vector meson. 

\subsubsection{The $3/2^{+}\to1/2^{+}$ case}

The corresponding decay widths are
\begin{align}
\Gamma({\cal B}\to{\cal B}^{\prime}P) & =\frac{|\lambda|^{2}f_{P}^{2}m^{2}|\vec{P}^{\prime}|}{32\pi M^{2}}(|H_{-\frac{1}{2},t}|^{2}+|H_{\frac{1}{2},t}|^{2}),\nonumber \\
\Gamma({\cal B}\to{\cal B}^{\prime}V) & =\frac{|\lambda|^{2}f_{V}^{2}m^{2}|\vec{P}^{\prime}|}{32\pi M^{2}}(|H_{-\frac{1}{2},-1}|^{2}+|H_{\frac{1}{2},-1}|^{2}+|H_{-\frac{1}{2},1}|^{2}+|H_{\frac{1}{2},1}|^{2}\nonumber \\
 & \qquad\qquad\qquad\qquad+|H_{-\frac{1}{2},0}|^{2}+|H_{\frac{1}{2},0}|^{2}),
\end{align}
where
\begin{equation}
\lambda\equiv\frac{G_{F}}{\sqrt{2}}\xi a_{1}
\end{equation}
with $\xi$ the product of the corresponding CKM matrix elements and
$a_{1}(\mu)\equiv C_{1}(\mu)+C_{2}(\mu)/3$, $|\vec{P}^{\prime}|$
is the magnitude of 3-momentum of ${\cal B}^{\prime}$ in the rest
frame of ${\cal B}$, and $m$ is the mass of the meson.

Our predictions for the nonleptonic decay widths of the $3/2\to1/2$
processes are given in Table \ref{Tab:non_31}. Some comments are
in order. 
\begin{itemize}
\item For the $c\to d/s$ processes, the decay width of ${\cal B}\to{\cal B}^{\prime}V$
is one order of magnitude larger than the corresponding decay width
of ${\cal B}\to{\cal B}^{\prime}P$, while for the $b\to u$ processes,
the ratio of the two is only around 2-3. 
\item Some words about so-called golden channels. Roughly speaking, the
decay mode ${\cal B}\to{\cal B}^{\prime}\pi$ is a candidate. This
situation is similar to that of $\Xi_{cc}^{++}$, for which the two
channels $\Xi_{cc}^{++}\to\Xi_{c}^{(\prime)+}\pi^{+}$ have been established
experimentally. However, if we choose, for example, the decay $\Omega_{ccc}^{++}\to\Xi_{cc}^{+}\pi^{+}$
as a golden channel to search for $\Omega_{ccc}^{++}$ , we should
discover $\Xi_{cc}^{+}$ first. Therefore, one may want to know the
decay width of $\Omega_{ccc}^{++}\to\Xi_{cc}^{*+}\pi^{+}$, where
the spin-3/2 $\Xi_{cc}^{*+}$ further decays into other charged final
states. These $3/2\to3/2$ processes are left to our future works.
\end{itemize}

\begin{table}
\caption{Decay widths for the nonleptonic decays of the $3/2\to1/2$ processes.}
\label{Tab:non_31}%
\begin{tabular}{l|c|l|c}
\hline 
Channel & $\Gamma/\text{~GeV}$ & Channel & $\Gamma/\text{~GeV}$\tabularnewline
\hline 
$\Omega_{ccc}^{++}\to\Xi_{cc}^{+}\pi^{+}$ & $1.39\times10^{-15}$ & $\Omega_{ccc}^{++}\to\Xi_{cc}^{+}\rho^{+}$ & $2.46\times10^{-14}$\tabularnewline
$\Omega_{ccc}^{++}\to\Xi_{cc}^{+}K^{+}$ & $1.41\times10^{-16}$ & $\Omega_{ccc}^{++}\to\Xi_{cc}^{+}K^{*+}$ & $1.65\times10^{-15}$\tabularnewline
\hline 
$\Omega_{bcc}^{\prime+}\to\Xi_{bc}^{\prime0}\pi^{+}$ & $1.05\times10^{-15}$ & $\Omega_{bcc}^{\prime+}\to\Xi_{bc}^{\prime0}\rho^{+}$ & $1.66\times10^{-14}$\tabularnewline
$\Omega_{bcc}^{\prime+}\to\Xi_{bc}^{\prime0}K^{+}$ & $9.71\times10^{-17}$ & $\Omega_{bcc}^{\prime+}\to\Xi_{bc}^{\prime0}K^{*+}$ & $1.04\times10^{-15}$\tabularnewline
\hline 
$\Omega_{bbc}^{\prime0}\to\Xi_{bb}^{-}\pi^{+}$ & $6.22\times10^{-16}$ & $\Omega_{bbc}^{\prime0}\to\Xi_{bb}^{-}\rho^{+}$ & $1.15\times10^{-14}$\tabularnewline
$\Omega_{bbc}^{\prime0}\to\Xi_{bb}^{-}K^{+}$ & $5.98\times10^{-17}$ & $\Omega_{bbc}^{\prime0}\to\Xi_{bb}^{-}K^{*+}$ & $7.27\times10^{-16}$\tabularnewline
\hline 
$\Omega_{ccc}^{++}\to\Omega_{cc}^{+}\pi^{+}$ & $2.91\times10^{-14}$ & $\Omega_{ccc}^{++}\to\Omega_{cc}^{+}\rho^{+}$ & $4.46\times10^{-13}$\tabularnewline
$\Omega_{ccc}^{++}\to\Omega_{cc}^{+}K^{+}$ & $2.62\times10^{-15}$ & $\Omega_{ccc}^{++}\to\Omega_{cc}^{+}K^{*+}$ & $2.70\times10^{-14}$\tabularnewline
\hline 
$\Omega_{bcc}^{\prime+}\to\Omega_{bc}^{\prime0}\pi^{+}$ & $2.10\times10^{-14}$ & $\Omega_{bcc}^{\prime+}\to\Omega_{bc}^{\prime0}\rho^{+}$ & $3.27\times10^{-13}$\tabularnewline
$\Omega_{bcc}^{\prime+}\to\Omega_{bc}^{\prime0}K^{+}$ & $1.84\times10^{-15}$ & $\Omega_{bcc}^{\prime+}\to\Omega_{bc}^{\prime0}K^{*+}$ & $1.88\times10^{-14}$\tabularnewline
\hline 
$\Omega_{bbc}^{\prime0}\to\Omega_{bb}^{-}\pi^{+}$ & $1.27\times10^{-14}$ & $\Omega_{bbc}^{\prime0}\to\Omega_{bb}^{-}\rho^{+}$ & $1.72\times10^{-13}$\tabularnewline
$\Omega_{bbc}^{\prime0}\to\Omega_{bb}^{-}K^{+}$ & $9.87\times10^{-16}$ & $\Omega_{bbc}^{\prime0}\to\Omega_{bb}^{-}K^{*+}$ & $7.89\times10^{-15}$\tabularnewline
\hline 
$\Omega_{bcc}^{\prime+}\to\Xi_{cc}^{++}\pi^{-}$ & $3.18\times10^{-20}$ & $\Omega_{bcc}^{\prime+}\to\Xi_{cc}^{++}\rho^{-}$ & $1.03\times10^{-19}$\tabularnewline
$\Omega_{bcc}^{\prime+}\to\Xi_{cc}^{++}K^{-}$ & $2.64\times10^{-21}$ & $\Omega_{bcc}^{\prime+}\to\Xi_{cc}^{++}K^{*-}$ & $5.55\times10^{-21}$\tabularnewline
$\Omega_{bcc}^{\prime+}\to\Xi_{cc}^{++}D^{-}$ & $5.15\times10^{-21}$ & $\Omega_{bcc}^{\prime+}\to\Xi_{cc}^{++}D^{*-}$ & $1.14\times10^{-20}$\tabularnewline
$\Omega_{bcc}^{\prime+}\to\Xi_{cc}^{++}D_{s}^{-}$ & $1.38\times10^{-19}$ & $\Omega_{bcc}^{\prime+}\to\Xi_{cc}^{++}D_{s}^{*-}$ & $2.86\times10^{-19}$\tabularnewline
\hline 
$\Omega_{bbc}^{\prime0}\to\Xi_{bc}^{\prime+}\pi^{-}$ & $1.70\times10^{-20}$ & $\Omega_{bbc}^{\prime0}\to\Xi_{bc}^{\prime+}\rho^{-}$ & $5.73\times10^{-20}$\tabularnewline
$\Omega_{bbc}^{\prime0}\to\Xi_{bc}^{\prime+}K^{-}$ & $1.42\times10^{-21}$ & $\Omega_{bbc}^{\prime0}\to\Xi_{bc}^{\prime+}K^{*-}$ & $3.13\times10^{-21}$\tabularnewline
$\Omega_{bbc}^{\prime0}\to\Xi_{bc}^{\prime+}D^{-}$ & $3.06\times10^{-21}$ & $\Omega_{bbc}^{\prime0}\to\Xi_{bc}^{\prime+}D^{*-}$ & $7.17\times10^{-21}$\tabularnewline
$\Omega_{bbc}^{\prime0}\to\Xi_{bc}^{\prime+}D_{s}^{-}$ & $8.26\times10^{-20}$ & $\Omega_{bbc}^{\prime0}\to\Xi_{bc}^{\prime+}D_{s}^{*-}$ & $1.81\times10^{-19}$\tabularnewline
\hline 
$\Omega_{bbb}^{-}\to\Xi_{bb}^{0}\pi^{-}$ & $1.15\times10^{-19}$ & $\Omega_{bbb}^{-}\to\Xi_{bb}^{0}\rho^{-}$ & $3.92\times10^{-19}$\tabularnewline
$\Omega_{bbb}^{-}\to\Xi_{bb}^{0}K^{-}$ & $9.60\times10^{-21}$ & $\Omega_{bbb}^{-}\to\Xi_{bb}^{0}K^{*-}$ & $2.15\times10^{-20}$\tabularnewline
$\Omega_{bbb}^{-}\to\Xi_{bb}^{0}D^{-}$ & $2.14\times10^{-20}$ & $\Omega_{bbb}^{-}\to\Xi_{bb}^{0}D^{*-}$ & $5.15\times10^{-20}$\tabularnewline
$\Omega_{bbb}^{-}\to\Xi_{bb}^{0}D_{s}^{-}$ & $5.79\times10^{-19}$ & $\Omega_{bbb}^{-}\to\Xi_{bb}^{0}D_{s}^{*-}$ & $1.30\times10^{-18}$\tabularnewline
\hline 
\end{tabular}
\end{table}

\subsubsection{The $1/2^{+}\to1/2^{+}$ case}

The corresponding decay widths are
\begin{align}
\Gamma({\cal B}\to{\cal B}^{\prime}P) & =\frac{|\lambda|^{2}f_{P}^{2}m^{2}|\vec{P}^{\prime}|}{16\pi M^{2}}(|H_{-\frac{1}{2},t}|^{2}+|H_{\frac{1}{2},t}|^{2}),\nonumber \\
\Gamma({\cal B}\to{\cal B}^{\prime}V/A) & =\frac{|\lambda|^{2}f_{V/A}^{2}m^{2}|\vec{P}^{\prime}|}{16\pi M^{2}}(|H_{-\frac{1}{2},-1}|^{2}+|H_{\frac{1}{2},1}|^{2}+|H_{-\frac{1}{2},0}|^{2}+|H_{\frac{1}{2},0}|^{2}).
\end{align}

Our predictions for the nonleptonic decay widths of the $1/2\to1/2$
processes are given in Tables \ref{Tab:non_11S} and \ref{Tab:non_11A}.
Some comments are in order. 
\begin{itemize}
\item We usually consider the spin-1/2 $\Omega_{bcc}^{+}$ ($\Omega_{bbc}^{0}$)
as the ground state and the spin-3/2 $\Omega_{bcc}^{\prime+}$ ($\Omega_{bbc}^{\prime0}$)
as an excited state. In experiments, the experimentalists
should prioritize searching for the former. 
\item It is interesting to compare the decay widths of these three decays:
$\Omega_{bcc}^{\prime+}\to\Omega_{bc}^{\prime0}\pi^{+}$, $\Omega_{bcc}^{+}\to\Omega_{bc}^{0}\pi^{+}$
and $\Omega_{bcc}^{+}\to\Omega_{bc}^{\prime0}\pi^{+}$, and we find: $\Gamma(\Omega_{bcc}^{\prime+}\to\Omega_{bc}^{\prime0}\pi^{+})<\Gamma(\Omega_{bcc}^{+}\to\Omega_{bc}^{\prime0}\pi^{+})<\Gamma(\Omega_{bcc}^{+}\to\Omega_{bc}^{0}\pi^{+})$.
\end{itemize}

\begin{table}
\caption{Decay widths for the nonleptonic decays of the $1/2\to1/2$ processes
with a scalar spectator diquark.}
\label{Tab:non_11S}%
\begin{tabular}{l|c|l|c}
\hline 
Channel & $\Gamma/\text{~GeV}$ & Channel & $\Gamma/\text{~GeV}$\tabularnewline
\hline 
$\Omega_{bcc}^{+}\to\Xi_{bc}^{0}\pi^{+}$ & $5.48\times10^{-15}$ & $\Omega_{bcc}^{+}\to\Xi_{bc}^{0}\rho^{+}$ & $3.17\times10^{-14}$\tabularnewline
$\Omega_{bcc}^{+}\to\Xi_{bc}^{0}K^{+}$ & $4.95\times10^{-16}$ & $\Omega_{bcc}^{+}\to\Xi_{bc}^{0}K^{*+}$ & $1.71\times10^{-15}$\tabularnewline
\hline 
$\Omega_{bcc}^{+}\to\Omega_{bc}^{0}\pi^{+}$ & $1.05\times10^{-13}$ & $\Omega_{bcc}^{+}\to\Omega_{bc}^{0}\rho^{+}$ & $5.86\times10^{-13}$\tabularnewline
$\Omega_{bcc}^{+}\to\Omega_{bc}^{0}K^{+}$ & $9.32\times10^{-15}$ & $\Omega_{bcc}^{+}\to\Omega_{bc}^{0}K^{*+}$ & $2.89\times10^{-14}$\tabularnewline
\hline 
$\Omega_{bbc}^{0}\to\Xi_{bc}^{+}\pi^{-}$ & $6.49\times10^{-20}$ & $\Omega_{bbc}^{0}\to\Xi_{bc}^{+}\rho^{-}$ & $1.96\times10^{-19}$\tabularnewline
$\Omega_{bbc}^{0}\to\Xi_{bc}^{+}K^{-}$ & $5.38\times10^{-21}$ & $\Omega_{bbc}^{0}\to\Xi_{bc}^{+}K^{*-}$ & $1.04\times10^{-20}$\tabularnewline
$\Omega_{bbc}^{0}\to\Xi_{bc}^{+}D^{-}$ & $1.07\times10^{-20}$ & $\Omega_{bbc}^{0}\to\Xi_{bc}^{+}D^{*-}$ & $1.74\times10^{-20}$\tabularnewline
$\Omega_{bbc}^{0}\to\Xi_{bc}^{+}D_{s}^{-}$ & $2.89\times10^{-19}$ & $\Omega_{bbc}^{0}\to\Xi_{bc}^{+}D_{s}^{*-}$ & $4.27\times10^{-19}$\tabularnewline
\hline 
\end{tabular}
\end{table}

\begin{table}
\caption{Decay widths for the nonleptonic decays of the $1/2\to1/2$ processes
with an axial-vector spectator diquark.}
\label{Tab:non_11A}%
\begin{tabular}{l|c|l|c}
\hline 
Channel & $\Gamma/\text{~GeV}$ & Channel & $\Gamma/\text{~GeV}$\tabularnewline
\hline 
$\Omega_{bcc}^{+}\to\Xi_{bc}^{\prime0}\pi^{+}$ & $1.31\times10^{-15}$ & $\Omega_{bcc}^{+}\to\Xi_{bc}^{\prime0}\rho^{+}$ & $2.58\times10^{-15}$\tabularnewline
$\Omega_{bcc}^{+}\to\Xi_{bc}^{\prime0}K^{+}$ & $1.18\times10^{-16}$ & $\Omega_{bcc}^{+}\to\Xi_{bc}^{\prime0}K^{*+}$ & $1.05\times10^{-16}$\tabularnewline
\hline 
$\Omega_{bbc}^{0}\to\Xi_{bb}^{-}\pi^{+}$ & $3.40\times10^{-15}$ & $\Omega_{bbc}^{0}\to\Xi_{bb}^{-}\rho^{+}$ & $7.40\times10^{-15}$\tabularnewline
$\Omega_{bbc}^{0}\to\Xi_{bb}^{-}K^{+}$ & $3.18\times10^{-16}$ & $\Omega_{bbc}^{0}\to\Xi_{bb}^{-}K^{*+}$ & $3.05\times10^{-16}$\tabularnewline
\hline 
$\Omega_{bcc}^{+}\to\Omega_{bc}^{\prime0}\pi^{+}$ & $2.46\times10^{-14}$ & $\Omega_{bcc}^{+}\to\Omega_{bc}^{\prime0}\rho^{+}$ & $4.30\times10^{-14}$\tabularnewline
$\Omega_{bcc}^{+}\to\Omega_{bc}^{\prime0}K^{+}$ & $2.23\times10^{-15}$ & $\Omega_{bcc}^{+}\to\Omega_{bc}^{\prime0}K^{*+}$ & $1.56\times10^{-15}$\tabularnewline
\hline 
$\Omega_{bbc}^{0}\to\Omega_{bb}^{-}\pi^{+}$ & $5.92\times10^{-14}$ & $\Omega_{bbc}^{0}\to\Omega_{bb}^{-}\rho^{+}$ & $8.71\times10^{-14}$\tabularnewline
$\Omega_{bbc}^{0}\to\Omega_{bb}^{-}K^{+}$ & $5.31\times10^{-15}$ & $\Omega_{bbc}^{0}\to\Omega_{bb}^{-}K^{*+}$ & $2.09\times10^{-15}$\tabularnewline
\hline 
$\Omega_{bcc}^{+}\to\Xi_{cc}^{++}\pi^{-}$ & $1.05\times10^{-19}$ & $\Omega_{bcc}^{+}\to\Xi_{cc}^{++}\rho^{-}$ & $2.87\times10^{-19}$\tabularnewline
$\Omega_{bcc}^{+}\to\Xi_{cc}^{++}K^{-}$ & $8.61\times10^{-21}$ & $\Omega_{bcc}^{+}\to\Xi_{cc}^{++}K^{*-}$ & $1.47\times10^{-20}$\tabularnewline
$\Omega_{bcc}^{+}\to\Xi_{cc}^{++}D^{-}$ & $1.45\times10^{-20}$ & $\Omega_{bcc}^{+}\to\Xi_{cc}^{++}D^{*-}$ & $1.57\times10^{-20}$\tabularnewline
$\Omega_{bcc}^{+}\to\Xi_{cc}^{++}D_{s}^{-}$ & $3.82\times10^{-19}$ & $\Omega_{bcc}^{+}\to\Xi_{cc}^{++}D_{s}^{*-}$ & $3.67\times10^{-19}$\tabularnewline
\hline 
$\Omega_{bbc}^{0}\to\Xi_{bc}^{\prime+}\pi^{-}$ & $1.49\times10^{-20}$ & $\Omega_{bbc}^{0}\to\Xi_{bc}^{\prime+}\rho^{-}$ & $4.08\times10^{-20}$\tabularnewline
$\Omega_{bbc}^{0}\to\Xi_{bc}^{\prime+}K^{-}$ & $1.22\times10^{-21}$ & $\Omega_{bbc}^{0}\to\Xi_{bc}^{\prime+}K^{*-}$ & $2.10\times10^{-21}$\tabularnewline
$\Omega_{bbc}^{0}\to\Xi_{bc}^{\prime+}D^{-}$ & $2.17\times10^{-21}$ & $\Omega_{bbc}^{0}\to\Xi_{bc}^{\prime+}D^{*-}$ & $2.30\times10^{-21}$\tabularnewline
$\Omega_{bbc}^{0}\to\Xi_{bc}^{\prime+}D_{s}^{-}$ & $5.75\times10^{-20}$ & $\Omega_{bbc}^{0}\to\Xi_{bc}^{\prime+}D_{s}^{*-}$ & $5.40\times10^{-20}$\tabularnewline
\hline 
\end{tabular}
\end{table}

\subsection{Uncertainties}

The shape parameter is one of the most important parameters in the
light-front approach, and in this work we use the pole residue to
determine this parameter. The pole residue is taken from the calculation
of QCD sum rules, which includes a given error. The error of the pole
residue is transmitted to the shape parameter, the errors in the shape
parameters of the initial and final baryons are then transmitted to
the form factors, and then to the semileptonic and nonleptonic decay
widths.

In the following, we will continue to consider these three typical
processes: $\Omega_{bcc}^{\prime+}\to\Omega_{bc}^{\prime0}$, $\Omega_{bcc}^{+}\to\Omega_{bc}^{0}$,
and $\Omega_{bcc}^{+}\to\Omega_{bc}^{\prime0}$. We tune the shape
parameters of initial and final baryons to the upper limits, thereby
obtaining the upper limits of the form factors and decay widths in
Table~\ref{Tab:uncertainties}. As a comparison, we also copy the
corresponding central values in this table. It can be seen that, the
uncertainties of the shape parameters of the initial and final baryons
result in about 15\% uncertainties for the semileptonic decay widths
and about 10\% uncertainties for the nonleptonic decay widths.

\begin{table}
\caption{Central values and upper limits for the form factors and decay widths
(in units of $10^{-14}$ GeV) of the three typical processes.}
\label{Tab:uncertainties} 

\begin{tabular}{l|c|c|c|c|c|c|c|c}
\hline 
$\Omega_{bcc}^{\prime+}\to\Omega_{bc}^{\prime0}$ & $f_{1}^{V}(0)$  & $f_{2}^{V}(0)$  & $f_{3}^{V}(0)$  & $f_{4}^{V}(0)$  & $f_{1}^{A}(0)$  & $f_{2}^{A}(0)$  & $f_{3}^{A}(0)$  & $f_{4}^{A}(0)$ \tabularnewline
\hline 
Central value  & $-2.431$ & $-0.315$ & $-2.072$ & $4.854$ & $23.062$ & $-37.360$ & $14.200$ & $1.123$\tabularnewline
Upper limit & $-2.500$  & $0.204$  & $-2.652$  & $4.983$ & $27.249$ & $-41.894$ & $14.690$ & $1.213$\tabularnewline
\hline 
\end{tabular}

\begin{tabular}{l|c|c|c|c|c|c}
\hline 
$\Omega_{bcc}^{+}\to\Omega_{bc}^{0}$ & $f_{1}(0)$ & $f_{2}(0)$ & $f_{3}(0)$ & $g_{1}(0)$ & $g_{2}(0)$ & $g_{3}(0)$\tabularnewline
\hline 
Central value & $-0.630$ & $3.269$ & $-1.500$ & $-0.445$ & $-2.479$ & $31.279$\tabularnewline
Upper limit & $-0.666$ & $3.833$ & $-1.852$ & $-0.458$ & $-2.814$ & $34.176$\tabularnewline
\hline 
\end{tabular}

\begin{tabular}{l|c|c|c|c|c|c}
\hline 
$\Omega_{bcc}^{+}\to\Omega_{bc}^{\prime0}$ & $f_{1}(0)$ & $f_{2}(0)$ & $f_{3}(0)$ & $g_{1}(0)$ & $g_{2}(0)$ & $g_{3}(0)$\tabularnewline
\hline 
Central value & $-0.363$ & $0.361$ & $-0.769$ & $0.086$ & $0.476$ & $-6.001$\tabularnewline
Upper limit & $-0.384$ & $0.647$ & $-0.968$ & $0.088$ & $0.540$ & $-6.557$\tabularnewline
\hline 
\end{tabular}

\begin{tabular}{l|c|c|c}
\hline 
Decay width & $\Gamma(\Omega_{bcc}^{\prime+}\to\Omega_{bc}^{\prime0}e^{+}\nu_{e})$ & $\Gamma(\Omega_{bcc}^{+}\to\Omega_{bc}^{0}e^{+}\nu_{e})$ & $\Gamma(\Omega_{bcc}^{+}\to\Omega_{bc}^{\prime0}e^{+}\nu_{e})$\tabularnewline
\hline 
Central value & $8.44$ & $15.7$ & $1.45$\tabularnewline
Upper limit & $9.66$ & $18.0$ & $1.66$\tabularnewline
\hline 
\end{tabular}

\begin{tabular}{l|c|c|c}
\hline 
Decay width & $\Gamma(\Omega_{bcc}^{\prime+}\to\Omega_{bc}^{\prime0}\pi^{+})$ & $\Gamma(\Omega_{bcc}^{+}\to\Omega_{bc}^{0}\pi^{+})$ & $\Gamma(\Omega_{bcc}^{+}\to\Omega_{bc}^{\prime0}\pi^{+})$\tabularnewline
\hline 
Central value & $2.10$ & $10.5$ & $2.46$\tabularnewline
Upper limit & $2.24$ & $11.6$ & $2.75$\tabularnewline
\hline 
\end{tabular}
\end{table}

\subsection{Comparison}

\label{subsec:Comparison}

Recently, Refs.~\cite{Wang:2022ias} and \cite{Lu:2024bqw} investigated
the weak decays of $\Omega_{ccc}^{++}$ and $\Omega_{bbb}^{-}$ in
the light-front approach, where the diquark picture and three-quark
picture are respectively adopted. In Table \ref{Tab:comparison},
our predicted decay widths are compared with those in these two works.
It can be seen that, there exist large differences. Some comments
are in order.
\begin{itemize}
\item The key parameters -- the shape parameters, in this work, are determined
using the corresponding pole residues of baryons, while Refs.~\cite{Wang:2022ias}
and \cite{Lu:2024bqw} selected these parameters based on experience.
The uncertainties in the shape parameters are the main source of error. 
\item For $\Gamma(\Omega_{ccc}^{++}\to\Xi_{cc}^{+}e^{+}\nu_{e})$ and $\Gamma(\Omega_{ccc}^{++}\to\Omega_{cc}^{+}e^{+}\nu_{e})$,
our results are several times larger than those in Refs.~\cite{Wang:2022ias}
and \cite{Lu:2024bqw}. This is due to the fact that we have adopted
larger shape parameters in this work. For example, in Ref.~\cite{Wang:2022ias},
the following values are adopted for the shape parameters
\begin{equation}
\beta_{c\{cc\}}=0.553,\quad\beta_{d\{cc\}}=0.370,\quad\beta_{s\{cc\}}=0.435,
\end{equation}
while in this work, they are respectively determined as
\begin{equation}
\beta_{\Omega_{ccc}}=0.704,\quad\beta_{\Xi_{cc}}=0.583,\quad\beta_{\Omega_{cc}}=0.600.
\end{equation}
\item For $\Gamma(\Omega_{bbb}^{-}\to\Xi_{bb}^{0}e^{-}\bar{\nu}_{e})$ and
$\Gamma(\Omega_{bbb}^{-}\to\Xi_{bb}^{0}\pi^{-})$, our results are
much smaller than those in Ref. \cite{Lu:2024bqw}. This is likely due
to the significant difference in the bottom quark mass used in Ref.
\cite{Lu:2024bqw} and that adopted in this work. In Ref. \cite{Lu:2024bqw},
$m_{b}=4.4$ GeV, while in this work, $m_{b}=4.8$ GeV. In addition,
Ref. \cite{Lu:2024bqw} adopted larger shape parameters. 
\end{itemize}
\begin{table}
\caption{Our predicted decay widths (in units of $10^{-14}$ GeV) are compared
with those in the literature.}
\label{Tab:comparison} %
\begin{tabular}{l|c|c|c|c}
\hline 
 & This work  & Ref.~\cite{Wang:2022ias} & Ref.~\cite{Lu:2024bqw}, Case I & Ref.~\cite{Lu:2024bqw}, Case II\tabularnewline
\hline 
$\Gamma(\Omega_{ccc}^{++}\to\Xi_{cc}^{+}e^{+}\nu_{e})$  & $1.11$  & $0.32$ & $0.307$ & $0.359$\tabularnewline
$\Gamma(\Omega_{ccc}^{++}\to\Omega_{cc}^{+}e^{+}\nu_{e})$  & $13.1$  & $4.95$ & $4.78$ & $5.51$\tabularnewline
$\Gamma(\Omega_{bbb}^{-}\to\Xi_{bb}^{0}e^{-}\bar{\nu}_{e})$  & $1.41\times10^{-3}$ & - - & $6.03\times10^{-3}$ & $10.1\times10^{-3}$\tabularnewline
\hline
$\Gamma(\Omega_{ccc}^{++}\to\Xi_{cc}^{+}\pi^{+})$  & $0.139$  & $0.160$ & $0.280$ & $0.362$\tabularnewline
$\Gamma(\Omega_{ccc}^{++}\to\Omega_{cc}^{+}\pi^{+})$  & $2.91$  & $7.32$ & $6.40$ & $8.03$\tabularnewline
$\Gamma(\Omega_{bbb}^{-}\to\Xi_{bb}^{0}\pi^{-})$  & $1.15\times10^{-5}$ & - - & $12.9\times10^{-5}$ & $17.8\times10^{-5}$\tabularnewline
\hline 
\end{tabular}
\end{table}

\section{Conclusions}

The recent progress in the study of doubly heavy baryons gives us
confidence to believe that we may not be far from the discovery of
triply heavy baryons. However, considering that there is still a lack
of one comprehensive quantitative analysis on the weak decays of triply
heavy baryons, this work and the forthcoming ones aim to fill this
gap.

Of course, we should first search for the ground-state triply heavy
baryons in experiments, which include: the spin-3/2 $\Omega_{ccc}^{++}$
and $\Omega_{bbb}^{-}$, and the spin-1/2 $\Omega_{bcc}^{+}$ and
$\Omega_{bbc}^{0}$. Therefore, in this work, we investigate the weak
decays of triply heavy baryons for the $3/2\to1/2$ case and $1/2\to1/2$
case. We first obtain the form factors using the light-front quark
model in the three-quark picture, and then apply them to arrive at
some phenomenological predictions, including the decay widths of semileptonic
decays and nonleptonic decays. Our results are expected to be helpful
for the experimental search for triply heavy baryons.

The road to searching for triply heavy baryons is long and arduous,
but the journey of a thousand miles begins with a single step. As
far as we are concerned, the following efforts can be made: 
\begin{itemize}
\item As mentioned above, the $3/2\to3/2$ processes may have
larger decay widths for the spin-3/2 $\Omega_{ccc}^{++}$ and $\Omega_{bbb}^{-}$.
In addition, for the sake of theoretical completeness, the $1/2\to3/2$
processes are also worth studying. In Ref.~\cite{Zhao:2018mrg},
we investigated the weak decays of doubly heavy baryons for the $1/2\to3/2$
case, and found that the decay widths are approximately one order
of magnitude smaller than those of the $1/2\to1/2$ case. One may
expect a similar pattern to hold for the case of triply heavy baryons. 
\item There is another type of processes that we deliberately avoid because
they do not have priority in experiments, but are also worth studying
for the sake of theoretical completeness, that is, the $b\to c$ processes.
\item As can be seen in Subsec.~\ref{subsec:Comparison}, despite all adopting
the light-front approach, different literatures still yield significantly
different results. Other methods, such as QCD sum rules and lattice
QCD, are needed to clarify these things. 
\item In order to search for triply heavy baryons, lifetime is another important
reference. Ref. \cite{Wang:2018utj} roughly estimated the lifetimes
of triply heavy baryons, however, the contribution of the four-quark
operators may play a significant role. We intend to make some efforts
in this direction in the future. 
\end{itemize}

\section*{Acknowledgements}

The authors are grateful to Profs.~Qin Chang, Chun-Khiang Chua, Run-Hui
Li, Wei Wang and Zhi-Peng Xing for valuable discussions. This work
is supported in part by National Natural Science Foundation of China
under Grant No.~12065020.
	
\appendix

\section{The normalization of the momentum wavefunction}

\label{app:normalization_momentum_wf}

In this appendix, we will demonstrate the normalization of the momentum
wavefunction, which requires:
\begin{equation}
\int\frac{dx_{1}d^{2}k_{1\perp}}{2(2\pi)^{3}}\frac{dx_{2}d^{2}k_{2\perp}}{2(2\pi)^{3}}\frac{dx_{3}d^{2}k_{3\perp}}{2(2\pi)^{3}}2(2\pi)^{3}\delta(1-\sum x_{i})\delta^{2}(\sum k_{i\perp})\frac{e_{1}e_{2}e_{3}}{x_{1}x_{2}x_{3}M_{0}}\varphi_{1}^{2}\varphi_{23}^{2}=1,\label{eq:normalization_momentum_wf}
\end{equation}
where we have denoted
\begin{equation}
\varphi_{1}=\varphi(\vec{k}_{1},\beta_{1}),\quad\varphi_{23}=\varphi(\frac{\vec{k}_{2}-\vec{k}_{3}}{2},\beta_{23}),
\end{equation}
and the defination of $\varphi$ can be found in Eq. (\ref{eq:momentum_wf_old}).

We rewrite the left-hand side of Eq. (\ref{eq:normalization_momentum_wf}):
\begin{align}
{\rm LHS} & =\int\frac{dx_{2}d^{2}k_{2\perp}}{2(2\pi)^{3}}\frac{dx_{3}d^{2}k_{3\perp}}{2(2\pi)^{3}}\frac{e_{1}e_{2}e_{3}}{x_{1}x_{2}x_{3}M_{0}}\varphi_{1}^{2}\varphi_{23}^{2}\nonumber \\
 & =\int\frac{dk_{2z}d^{2}k_{2\perp}}{2(2\pi)^{3}}\frac{dk_{3z}d^{2}k_{3\perp}}{2(2\pi)^{3}}\varphi^{2}(\vec{k}_{1},\beta_{1})\varphi^{2}(\frac{\vec{k}_{2}-\vec{k}_{3}}{2},\beta_{23})\nonumber \\
 & =\int\frac{d^{3}k_{2}}{2(2\pi)^{3}}\frac{d^{3}k_{3}}{2(2\pi)^{3}}\varphi^{2}(\vec{k}_{2}+\vec{k}_{3},\beta_{1})\varphi^{2}(\frac{\vec{k}_{2}-\vec{k}_{3}}{2},\beta_{23})\nonumber \\
 & =\int\frac{d^{3}k_{23}}{2(2\pi)^{3}}\frac{d^{3}l_{23}}{2(2\pi)^{3}}\varphi^{2}(\vec{k}_{23},\beta_{1})\varphi^{2}(\vec{l}_{23},\beta_{23})\nonumber \\
 & =1={\rm RHS}.
\end{align}
In the second step, we have used 
\begin{equation}
\left|\begin{array}{cc}
\frac{\partial k_{2z}}{\partial x_{2}} & \frac{\partial k_{2z}}{\partial x_{3}}\\
\frac{\partial k_{3z}}{\partial x_{2}} & \frac{\partial k_{3z}}{\partial x_{3}}
\end{array}\right|=\frac{e_{1}e_{2}e_{3}}{x_{1}x_{2}x_{3}M_{0}},
\end{equation}
in the fourth step, we define $\vec{k}_{23}\equiv\vec{k}_{2}+\vec{k}_{3}$
and $\vec{l}_{23}\equiv\vec{k}_{2}-\vec{k}_{3}/2$, and note that
$\int d^{3}k_{2}d^{3}k_{3}=\int d^{3}k_{23}d^{3}l_{23}$, and in the
last step,
\begin{equation}
\int\frac{d^{3}k}{2(2\pi)^{3}}\varphi^{2}(\vec{k},\beta)=1.
\end{equation}

In Eq. (\ref{eq:normalization_momentum_wf}), if we replace $\varphi_{1}^{2}\varphi_{23}^{2}$ with $\varphi_{123}^{2}$, where
\begin{equation}
\varphi_{123}\sim\exp\left(-\frac{\vec{k}_{1}^{2}+\vec{k}_{2}^{2}+\vec{k}_{3}^{2}}{2\beta^{2}}\right)=\exp\left(-\frac{\frac{3}{2}\vec{k}_{23}^{2}+2\vec{l}_{23}^{2}}{2\beta^{2}}\right),
\end{equation}
then one can easily obtain the normalization factor of $\varphi_{123}$,
see Eq. (\ref{eq:momentum_wf}).


\begin{thebibliography}{1}

\bibitem{LHCb:2017iph}
R.~Aaij \textit{et al.} [LHCb],
Phys. Rev. Lett. \textbf{119}, no.11, 112001 (2017)
doi:10.1103/PhysRevLett.119.112001
[arXiv:1707.01621 [hep-ex]].

\bibitem{LHCb:2018zpl}
R.~Aaij \textit{et al.} [LHCb],
Phys. Rev. Lett. \textbf{121}, no.5, 052002 (2018)
doi:10.1103/PhysRevLett.121.052002
[arXiv:1806.02744 [hep-ex]].

\bibitem{LHCb:2018pcs}
R.~Aaij \textit{et al.} [LHCb],
Phys. Rev. Lett. \textbf{121}, no.16, 162002 (2018)
doi:10.1103/PhysRevLett.121.162002
[arXiv:1807.01919 [hep-ex]].

\bibitem{LHCb:2022rpd}
R.~Aaij \textit{et al.} [LHCb],
JHEP \textbf{05}, 038 (2022)
doi:10.1007/JHEP05(2022)038
[arXiv:2202.05648 [hep-ex]].

\bibitem{Wang:2020avt}
Z.~G.~Wang,
AAPPS Bull. \textbf{31}, 5 (2021)
doi:10.1007/s43673-021-00006-3
[arXiv:2010.08939 [hep-ph]].

\bibitem{Huang:2021jxt}
F.~Huang, J.~Xu and X.~R.~Zhang,
Eur. Phys. J. C \textbf{81}, no.11, 976 (2021)
doi:10.1140/epjc/s10052-021-09729-x
[arXiv:2107.13958 [hep-ph]].

\bibitem{Wang:2022ias}
W.~Wang and Z.~P.~Xing,
Phys. Lett. B \textbf{834}, 137402 (2022)
doi:10.1016/j.physletb.2022.137402
[arXiv:2203.14446 [hep-ph]].

\bibitem{Lu:2024bqw}
F.~Lu, H.~W.~Ke and X.~H.~Liu,
Eur. Phys. J. C \textbf{84}, no.5, 452 (2024)
doi:10.1140/epjc/s10052-024-12732-7
[arXiv:2402.01334 [hep-ph]].

\bibitem{GomshiNobary:2004mq}
M.~A.~Gomshi Nobary and R.~Sepahvand,
Phys. Rev. D \textbf{71}, 034024 (2005)
doi:10.1103/PhysRevD.71.034024
[arXiv:hep-ph/0406148 [hep-ph]].

\bibitem{GomshiNobary:2005ur}
M.~A.~Gomshi Nobary and R.~Sepahvand,
Nucl. Phys. B \textbf{741}, 34-41 (2006)
doi:10.1016/j.nuclphysb.2006.01.043
[arXiv:hep-ph/0508115 [hep-ph]].

\bibitem{Jia:2006gw}
Y.~Jia,
JHEP \textbf{10}, 073 (2006)
doi:10.1088/1126-6708/2006/10/073
[arXiv:hep-ph/0607290 [hep-ph]].

\bibitem{GomshiNobary:2006tzy}
M.~A.~Gomshi Nobary and R.~Sepahvand,
eConf \textbf{C0605151}, 0010 (2006)
[arXiv:hep-ph/0609123 [hep-ph]].

\bibitem{Martynenko:2007je}
A.~P.~Martynenko,
Phys. Lett. B \textbf{663}, 317-321 (2008)
doi:10.1016/j.physletb.2008.04.030
[arXiv:0708.2033 [hep-ph]].

\bibitem{Patel:2008mv}
B.~Patel, A.~Majethiya and P.~C.~Vinodkumar,
Pramana \textbf{72}, 679-688 (2009)
doi:10.1007/s12043-009-0061-4
[arXiv:0808.2880 [hep-ph]].

\bibitem{Zhang:2009re}
J.~R.~Zhang and M.~Q.~Huang,
Phys. Lett. B \textbf{674}, 28-35 (2009)
doi:10.1016/j.physletb.2009.02.056
[arXiv:0902.3297 [hep-ph]].

\bibitem{Zheng:2010zzc}
W.~Zheng and H.~R.~Pang,
Mod. Phys. Lett. A \textbf{25}, 2077-2088 (2010)
doi:10.1142/S0217732310032962

\bibitem{Chen:2011mb}
Y.~Q.~Chen and S.~Z.~Wu,
JHEP \textbf{08}, 144 (2011)
[erratum: JHEP \textbf{09}, 089 (2011)]
doi:10.1007/JHEP08(2011)144
[arXiv:1106.0193 [hep-ph]].

\bibitem{Flynn:2011gf}
J.~M.~Flynn, E.~Hernandez and J.~Nieves,
Phys. Rev. D \textbf{85}, 014012 (2012)
doi:10.1103/PhysRevD.85.014012
[arXiv:1110.2962 [hep-ph]].

\bibitem{Llanes-Estrada:2011gwu}
F.~J.~Llanes-Estrada, O.~I.~Pavlova and R.~Williams,
Eur. Phys. J. C \textbf{72}, 2019 (2012)
doi:10.1140/epjc/s10052-012-2019-9
[arXiv:1111.7087 [hep-ph]].

\bibitem{Wang:2011ae}
Z.~G.~Wang,
Commun. Theor. Phys. \textbf{58}, 723-731 (2012)
doi:10.1088/0253-6102/58/5/17
[arXiv:1112.2274 [hep-ph]].

\bibitem{Albertus:2012isp}
C.~Albertus, J.~M.~Flynn, E.~Hernandez and J.~Nieves,
PoS \textbf{ConfinementX}, 146 (2012)
doi:10.22323/1.171.0146
[arXiv:1301.3024 [hep-ph]].

\bibitem{Aliev:2012tt}
T.~M.~Aliev, K.~Azizi and M.~Savci,
JHEP \textbf{04}, 042 (2013)
doi:10.1007/JHEP04(2013)042
[arXiv:1212.6065 [hep-ph]].

\bibitem{Aliev:2014lxa}
T.~M.~Aliev, K.~Azizi and M.~Savc\i{},
J. Phys. G \textbf{41}, 065003 (2014)
doi:10.1088/0954-3899/41/6/065003
[arXiv:1404.2091 [hep-ph]].

\bibitem{Azizi:2014jxa}
K.~Azizi, T.~M.~Aliev and M.~Savci,
J. Phys. Conf. Ser. \textbf{556}, no.1, 012016 (2014)
doi:10.1088/1742-6596/556/1/012016

\bibitem{Vijande:2015faa}
J.~Vijande, A.~Valcarce and H.~Garcilazo,
Phys. Rev. D \textbf{91}, no.5, 054011 (2015)
doi:10.1103/PhysRevD.91.054011
[arXiv:1507.03735 [hep-ph]].

\bibitem{Thakkar:2016cdn}
K.~Thakkar, A.~Majethiya and P.~C.~Vinodkumar,
DAE Symp. Nucl. Phys. \textbf{61}, 710-711 (2016)

\bibitem{Shah:2017jkr}
Z.~Shah and A.~K.~Rai,
Eur. Phys. J. A \textbf{53}, no.10, 195 (2017)
doi:10.1140/epja/i2017-12386-2

\bibitem{MoosaviNejad:2017rvi}
S.~M.~Moosavi Nejad,
Phys. Rev. D \textbf{96}, no.11, 114021 (2017)
doi:10.1103/PhysRevD.96.114021

\bibitem{Rai:2017hue}
A.~K.~Rai and Z.~Shah,
J. Phys. Conf. Ser. \textbf{934}, no.1, 012035 (2017)
doi:10.1088/1742-6596/934/1/012035

\bibitem{Bhavsar:2018tad}
T.~Bhavsar, M.~Shah and P.~C.~Vinodkumar,
DAE Symp. Nucl. Phys. \textbf{63}, 840-841 (2018)

\bibitem{Wang:2018utj}
W.~Wang and J.~Xu,
Phys. Rev. D \textbf{97}, no.9, 093007 (2018)
doi:10.1103/PhysRevD.97.093007
[arXiv:1803.01476 [hep-ph]].

\bibitem{Shah:2019jxp}
Z.~Shah and A.~K.~Rai,
EPJ Web Conf. \textbf{202}, 06001 (2019)
doi:10.1051/epjconf/201920206001

\bibitem{Yang:2019lsg}
G.~Yang, J.~Ping, P.~G.~Ortega and J.~Segovia,
Chin. Phys. C \textbf{44}, no.2, 023102 (2020)
doi:10.1088/1674-1137/44/2/023102
[arXiv:1904.10166 [hep-ph]].

\bibitem{Delpasand:2019xpk}
M.~Delpasand and S.~M.~Moosavi Nejad,
Phys. Rev. D \textbf{99}, no.11, 114028 (2019)
doi:10.1103/PhysRevD.99.114028

\bibitem{Alomayrah:2020qyw}
N.~Alomayrah and T.~Barakat,
Eur. Phys. J. A \textbf{56}, no.3, 76 (2020)
doi:10.1140/epja/s10050-020-00062-7

\bibitem{Tazimi:2021ywr}
N.~Tazimi and A.~Ghasempour,
Mod. Phys. Lett. A \textbf{36}, no.39, 2150270 (2021)
doi:10.1142/S0217732321502709

\bibitem{Ghasemi:2021wqo}
M.~Ghasemi and R.~Sepahvand,
Int. J. Theor. Phys. \textbf{60}, no.4, 1261-1274 (2021)
doi:10.1007/s10773-021-04752-w

\bibitem{Mutuk:2021zes}
H.~Mutuk and U.~\"Ozdem,
Eur. Phys. J. Plus \textbf{137}, no.4, 508 (2022)
doi:10.1140/epjp/s13360-022-02724-5
[arXiv:2107.04361 [hep-ph]].

\bibitem{Faustov:2021qqf}
R.~N.~Faustov and V.~O.~Galkin,
Phys. Rev. D \textbf{105}, no.1, 014013 (2022)
doi:10.1103/PhysRevD.105.014013
[arXiv:2111.07702 [hep-ph]].

\bibitem{Sun:2023noo}
X.~Y.~Sun, F.~W.~Zhang, Y.~J.~Shi and Z.~X.~Zhao,
Eur. Phys. J. C \textbf{83}, no.10, 961 (2023)
doi:10.1140/epjc/s10052-023-12042-4
[arXiv:2305.08050 [hep-ph]].

\bibitem{Jaus:1999zv}
W.~Jaus,
Phys. Rev. D \textbf{60}, 054026 (1999)
doi:10.1103/PhysRevD.60.054026

\bibitem{Jaus:1989au}
W.~Jaus,
Phys. Rev. D \textbf{41}, 3394 (1990)
doi:10.1103/PhysRevD.41.3394

\bibitem{Jaus:1991cy}
W.~Jaus,
Phys. Rev. D \textbf{44}, 2851-2859 (1991)
doi:10.1103/PhysRevD.44.2851

\bibitem{Cheng:1996if}
H.~Y.~Cheng, C.~Y.~Cheung and C.~W.~Hwang,
Phys. Rev. D \textbf{55}, 1559-1577 (1997)
doi:10.1103/PhysRevD.55.1559
[arXiv:hep-ph/9607332 [hep-ph]].

\bibitem{Cheng:2003sm}
H.~Y.~Cheng, C.~K.~Chua and C.~W.~Hwang,
Phys. Rev. D \textbf{69}, 074025 (2004)
doi:10.1103/PhysRevD.69.074025
[arXiv:hep-ph/0310359 [hep-ph]].

\bibitem{Cheng:2004yj}
H.~Y.~Cheng and C.~K.~Chua,
Phys. Rev. D \textbf{69}, 094007 (2004)
[erratum: Phys. Rev. D \textbf{81}, 059901 (2010)]
doi:10.1103/PhysRevD.69.094007
[arXiv:hep-ph/0401141 [hep-ph]].

\bibitem{Ke:2009ed}
H.~W.~Ke, X.~Q.~Li and Z.~T.~Wei,
Phys. Rev. D \textbf{80}, 074030 (2009)
doi:10.1103/PhysRevD.80.074030
[arXiv:0907.5465 [hep-ph]].

\bibitem{Ke:2009mn}
H.~W.~Ke, X.~Q.~Li and Z.~T.~Wei,
Eur. Phys. J. C \textbf{69}, 133-138 (2010)
doi:10.1140/epjc/s10052-010-1383-6
[arXiv:0912.4094 [hep-ph]].

\bibitem{Cheng:2009ms}
H.~Y.~Cheng and C.~K.~Chua,
Phys. Rev. D \textbf{81}, 114006 (2010)
[erratum: Phys. Rev. D \textbf{82}, 059904 (2010)]
doi:10.1103/PhysRevD.81.114006
[arXiv:0909.4627 [hep-ph]].

\bibitem{Lu:2007sg}
C.~D.~Lu, W.~Wang and Z.~T.~Wei,
Phys. Rev. D \textbf{76}, 014013 (2007)
doi:10.1103/PhysRevD.76.014013
[arXiv:hep-ph/0701265 [hep-ph]].

\bibitem{Wang:2007sxa}
W.~Wang, Y.~L.~Shen and C.~D.~Lu,
Eur. Phys. J. C \textbf{51}, 841-847 (2007)
doi:10.1140/epjc/s10052-007-0334-3
[arXiv:0704.2493 [hep-ph]].

\bibitem{Wang:2008xt}
W.~Wang, Y.~L.~Shen and C.~D.~Lu,
Phys. Rev. D \textbf{79}, 054012 (2009)
doi:10.1103/PhysRevD.79.054012
[arXiv:0811.3748 [hep-ph]].

\bibitem{Wang:2008ci}
W.~Wang and Y.~L.~Shen,
Phys. Rev. D \textbf{78}, 054002 (2008)
doi:10.1103/PhysRevD.78.054002

\bibitem{Wang:2009mi}
X.~X.~Wang, W.~Wang and C.~D.~Lu,
Phys. Rev. D \textbf{79}, 114018 (2009)
doi:10.1103/PhysRevD.79.114018
[arXiv:0901.1934 [hep-ph]].

\bibitem{Chen:2009qk}
C.~H.~Chen, Y.~L.~Shen and W.~Wang,
Phys. Lett. B \textbf{686}, 118-123 (2010)
doi:10.1016/j.physletb.2010.02.056
[arXiv:0911.2875 [hep-ph]].

\bibitem{Li:2010bb}
G.~Li, F.~l.~Shao and W.~Wang,
Phys. Rev. D \textbf{82}, 094031 (2010)
doi:10.1103/PhysRevD.82.094031
[arXiv:1008.3696 [hep-ph]].

\bibitem{Verma:2011yw}
R.~C.~Verma,
J. Phys. G \textbf{39}, 025005 (2012)
doi:10.1088/0954-3899/39/2/025005
[arXiv:1103.2973 [hep-ph]].

\bibitem{Shi:2016gqt}
Y.~J.~Shi, W.~Wang and Z.~X.~Zhao,
Eur. Phys. J. C \textbf{76}, no.10, 555 (2016)
doi:10.1140/epjc/s10052-016-4405-1
[arXiv:1607.00622 [hep-ph]].

\bibitem{Chang:2018aut}
Q.~Chang, X.~N.~Li, X.~Q.~Li and F.~Su,
Chin. Phys. C \textbf{42}, no.7, 073102 (2018)
doi:10.1088/1674-1137/42/7/073102
[arXiv:1805.00718 [hep-ph]].

\bibitem{Chang:2018zjq}
Q.~Chang, X.~N.~Li, X.~Q.~Li, F.~Su and Y.~D.~Yang,
Phys. Rev. D \textbf{98}, no.11, 114018 (2018)
doi:10.1103/PhysRevD.98.114018
[arXiv:1810.00296 [hep-ph]].

\bibitem{Chang:2019mmh}
Q.~Chang, X.~N.~Li and L.~T.~Wang,
Eur. Phys. J. C \textbf{79}, no.5, 422 (2019)
doi:10.1140/epjc/s10052-019-6949-3
[arXiv:1905.05098 [hep-ph]].

\bibitem{Chang:2019obq}
Q.~Chang, L.~T.~Wang and X.~N.~Li,
JHEP \textbf{12}, 102 (2019)
doi:10.1007/JHEP12(2019)102
[arXiv:1908.04677 [hep-ph]].

\bibitem{Chang:2020wvs}
Q.~Chang, X.~L.~Wang and L.~T.~Wang,
Chin. Phys. C \textbf{44}, no.8, 083105 (2020)
doi:10.1088/1674-1137/44/8/083105
[arXiv:2003.10833 [hep-ph]].

\bibitem{Chen:2021ywv}
L.~Chen, Y.~W.~Ren, L.~T.~Wang and Q.~Chang,
Eur. Phys. J. C \textbf{82}, no.5, 451 (2022)
doi:10.1140/epjc/s10052-022-10391-0
[arXiv:2112.08016 [hep-ph]].

\bibitem{Ke:2007tg}
H.~W.~Ke, X.~Q.~Li and Z.~T.~Wei,
Phys. Rev. D \textbf{77}, 014020 (2008)
doi:10.1103/PhysRevD.77.014020
[arXiv:0710.1927 [hep-ph]].

\bibitem{Wei:2009np}
Z.~T.~Wei, H.~W.~Ke and X.~Q.~Li,
Phys. Rev. D \textbf{80}, 094016 (2009)
doi:10.1103/PhysRevD.80.094016
[arXiv:0909.0100 [hep-ph]].

\bibitem{Ke:2012wa}
H.~W.~Ke, X.~H.~Yuan, X.~Q.~Li, Z.~T.~Wei and Y.~X.~Zhang,
Phys. Rev. D \textbf{86}, 114005 (2012)
doi:10.1103/PhysRevD.86.114005
[arXiv:1207.3477 [hep-ph]].

\bibitem{Hu:2020mxk}
X.~H.~Hu, R.~H.~Li and Z.~P.~Xing,
Eur. Phys. J. C \textbf{80}, no.4, 320 (2020)
doi:10.1140/epjc/s10052-020-7851-8
[arXiv:2001.06375 [hep-ph]].

\bibitem{Hsiao:2020gtc}
Y.~K.~Hsiao, L.~Yang, C.~C.~Lih and S.~Y.~Tsai,
Eur. Phys. J. C \textbf{80}, no.11, 1066 (2020)
doi:10.1140/epjc/s10052-020-08619-y
[arXiv:2009.12752 [hep-ph]].

\bibitem{Ke:2019smy}
H.~W.~Ke, N.~Hao and X.~Q.~Li,
Eur. Phys. J. C \textbf{79}, no.6, 540 (2019)
doi:10.1140/epjc/s10052-019-7048-1
[arXiv:1904.05705 [hep-ph]].

\bibitem{Ke:2019lcf}
H.~W.~Ke, F.~Lu, X.~H.~Liu and X.~Q.~Li,
Eur. Phys. J. C \textbf{80}, no.2, 140 (2020)
doi:10.1140/epjc/s10052-020-7699-y
[arXiv:1912.01435 [hep-ph]].

\bibitem{Ke:2021pxk}
H.~W.~Ke, Q.~Q.~Kang, X.~H.~Liu and X.~Q.~Li,
Chin. Phys. C \textbf{45}, no.11, 113103 (2021)
doi:10.1088/1674-1137/ac1c66
[arXiv:2106.07013 [hep-ph]].

\bibitem{Geng:2020fng}
C.~Q.~Geng, C.~C.~Lih, C.~W.~Liu and T.~H.~Tsai,
Phys. Rev. D \textbf{101}, no.9, 094017 (2020)
doi:10.1103/PhysRevD.101.094017
[arXiv:2002.10612 [hep-ph]].

\bibitem{Geng:2020gjh}
C.~Q.~Geng, C.~W.~Liu and T.~H.~Tsai,
Phys. Rev. D \textbf{103}, no.5, 054018 (2021)
doi:10.1103/PhysRevD.103.054018
[arXiv:2012.04147 [hep-ph]].

\bibitem{Geng:2021nkl}
C.~Q.~Geng, C.~W.~Liu and T.~H.~Tsai,
Phys. Lett. B \textbf{815}, 136125 (2021)
doi:10.1016/j.physletb.2021.136125
[arXiv:2102.01552 [hep-ph]].

\bibitem{Geng:2022xpn}
C.~Q.~Geng, C.~W.~Liu, Z.~Y.~Wei and J.~Zhang,
Phys. Rev. D \textbf{105}, no.7, 073007 (2022)
doi:10.1103/PhysRevD.105.073007
[arXiv:2202.06179 [hep-ph]].

\bibitem{Zhao:2023yuk}
Z.~X.~Zhao, F.~W.~Zhang, X.~H.~Hu and Y.~J.~Shi,
Phys. Rev. D \textbf{107}, no.11, 116025 (2023)
doi:10.1103/PhysRevD.107.116025
[arXiv:2304.07698 [hep-ph]].

\bibitem{Xing:2023jnr}
Z.~P.~Xing, Y.~J.~Shi, J.~Sun and Z.~X.~Zhao,
[arXiv:2312.17568 [hep-ph]].

\bibitem{Wang:2017mqp}
W.~Wang, F.~S.~Yu and Z.~X.~Zhao,
Eur. Phys. J. C \textbf{77}, no.11, 781 (2017)
doi:10.1140/epjc/s10052-017-5360-1
[arXiv:1707.02834 [hep-ph]].

\bibitem{Chua:2018lfa}
C.~K.~Chua,
Phys. Rev. D \textbf{99}, no.1, 014023 (2019)
doi:10.1103/PhysRevD.99.014023
[arXiv:1811.09265 [hep-ph]].

\bibitem{Brown:2014ena}
Z.~S.~Brown, W.~Detmold, S.~Meinel and K.~Orginos,
Phys. Rev. D \textbf{90}, no.9, 094507 (2014)
doi:10.1103/PhysRevD.90.094507
[arXiv:1409.0497 [hep-lat]].

\bibitem{Hu:2017dzi}
X.~H.~Hu, Y.~L.~Shen, W.~Wang and Z.~X.~Zhao,
Chin. Phys. C \textbf{42}, no.12, 123102 (2018)
doi:10.1088/1674-1137/42/12/123102
[arXiv:1711.10289 [hep-ph]].

\bibitem{Zhao:2018mrg}
Z.~X.~Zhao,
Eur. Phys. J. C \textbf{78}, no.9, 756 (2018)
doi:10.1140/epjc/s10052-018-6213-2
[arXiv:1805.10878 [hep-ph]].

\bibitem{Carrasco:2014poa}
N.~Carrasco, P.~Dimopoulos, R.~Frezzotti, P.~Lami, V.~Lubicz, F.~Nazzaro, E.~Picca, L.~Riggio, G.~C.~Rossi and F.~Sanfilippo, \textit{et al.}
Phys. Rev. D \textbf{91}, no.5, 054507 (2015)
doi:10.1103/PhysRevD.91.054507
[arXiv:1411.7908 [hep-lat]].

\bibitem{Buras:1998raa}
A.~J.~Buras,
[arXiv:hep-ph/9806471 [hep-ph]].

\bibitem{ParticleDataGroup:2022pth}
R.~L.~Workman \textit{et al.} [Particle Data Group],
PTEP \textbf{2022}, 083C01 (2022)
doi:10.1093/ptep/ptac097

\end{thebibliography}
\end{document}